\RequirePackage{fixltx2e}
\documentclass[12pt,a4paper]{iopart}
\pdfoutput=1
\usepackage[utf8]{inputenc}
\usepackage[unicode=true,bookmarksopen=true,colorlinks=true,urlcolor=blue,anchorcolor=blue,citecolor=blue,filecolor=blue,linkcolor=blue,menucolor=blue,linktocpage=true,pdfa=true,hypertexnames=false]{hyperref}

\makeatletter
\providecommand*\@nameundef[1]{\expandafter\let\csname #1\endcsname\@undefined}
\@nameundef{equation*}
\@nameundef{endequation*}
\makeatother
\usepackage{amsmath}
\usepackage{amssymb}
\usepackage{amsfonts}
\usepackage{mathtools}
\usepackage{lmodern}
\usepackage[T1]{fontenc}
\usepackage{bbm}
\DeclareMathAlphabet{\mathbfi}{OML}{cmm}{b}{it}

\let\originalleft\left
\let\originalright\right
\renewcommand{\left}{\mathopen{}\mathclose\bgroup\originalleft}
\renewcommand{\right}{\aftergroup\egroup\originalright}

\makeatletter
\newenvironment{equations}[1][]{\subequations\ifx\relax#1\relax\else\label{#1}\fi\align\ignorespaces}{\endalign\ignorespacesafterend\endsubequations}
\def\@spliteq#1{\begin{equation}\begin{split}#1\end{split}\end{equation}}
\def\splitequation{\collect@body\@spliteq}

\makeatother

\renewcommand{\vec}[1]{{\ifnum9<1#1\mathbf{#1}\else\ifcat\noexpand#1\relax\boldsymbol{#1}\else\mathbfi{#1}\fi\fi}}
\newcommand{\mathe}{\mathrm{e}}
\newcommand{\mathi}{\mathrm{i}}
\let\oldre\Re
\let\oldim\Im
\renewcommand{\Re}{\oldre\mathfrak{e}\,}
\renewcommand{\Im}{\oldim\mathfrak{m}\,}
\newcommand{\total}{\mathop{}\!\mathrm{d}}
\newcommand{\laplace}{\mathop{}\!\bigtriangleup}
\newcommand{\abs}[1]{{\left\lvert{#1}\right\rvert}}

\newcommand{\eqend}[1]{\,#1}
\newcommand{\bigo}[1]{\mathcal{O}\left({#1}\right)}

\newcommand{\expect}[1]{\left\langle{#1}\right\rangle}

\newcommand{\hankel}[1]{\mathop{}\!\mathrm{H}^{(#1)}}

\usepackage[numbers,sort&compress]{natbib}
\allowdisplaybreaks

\begin{document}

\title{One-loop quantum gravitational backreaction on the local Hubble rate}

\author{Markus B. Fr{\"o}b}
\address{Institut f{\"u}r Theoretische Physik, Universit{\"a}t Leipzig,\\ Br{\"u}derstra{\ss}e 16, 04103 Leipzig, Germany\footnote{Previous address: Department of Mathematics, University of York, Heslington, York, YO10 5DD, United Kingdom}}
\ead{\href{mailto:mfroeb@itp.uni-leipzig.de}{mfroeb@itp.uni-leipzig.de}}

\begin{abstract}
We determine corrections to the Hubble rate due to graviton loops in a cosmological background spacetime of constant deceleration parameter. The corrections are gauge-invariant, based on a recent proposal for all-order gauge-invariant observables in perturbative quantum gravity. We find explicit expressions for the cases of matter- and radiation-dominated eras and slow-roll inflation with vanishing second slow-roll parameter. Interestingly, in the latter case the corrections can be described by a quantum-corrected first slow-roll parameter, which brings the spacetime closer to de~Sitter space.

\end{abstract}

\submitto{CQG}

\maketitle

\section{Introduction}
\label{sec_introduction}

Backreaction effects in cosmology, and more specifically backreaction effects from quantum fluctuations in primordial inflation, have been studied by many authors, with various partly conflicting results and conclusions, see, e.g.~\cite{mukhanovabramobrandenberger1997,unruh1998,abramowoodard1999a,abramowoodard1999b,abramowoodard2002,geshnizjanibrandenberger2002,geshnizjanibrandenberger2005,geshnizjaniafshordi2005,losicunruh2005,marozzivacca2012,marozzivaccabrandenberger2012,miaotsamiswoodard2017} and references therein. The calculations can be done perturbatively, by treating quantum gravity (which is known to be non-renormalisable as a quantum field theory) as an effective theory~\cite{burgess2004}, considering metric fluctuations around a classical inflationary background. Unambiguous predictions can then be made at scales well below the Planck scale, which includes the power spectra of tree-level scalar metric fluctuations measured from the cosmic microwave background~\cite{planck2015a,planck2015b,planck2015c}, our only evidence for quantum gravity to date. A main obstacle for including the effects of graviton loops has been the identification of suitable observables. In perturbative quantum gravity, diffeomorphism invariance translates into a gauge symmetry for the graviton, and in contrast to theories where the gauge symmetry is an internal symmetry, it is known that no local gauge-invariant observables (defined at a point) exist in quantum gravity~\cite{torre1993,giddingsmarolfhartle2006,khavkine2015}. To identify suitable (necessarily non-local) observables, we propose the following two criteria:
\begin{enumerate}
\item \textbf{Gauge invariance:} Since the gauge symmetry of perturbative quantum gravity, coming from the general coordinate invariance of the underlying gravity theory, is not a physical symmetry but only a redundancy in the description, physical quantities must be unchanged under a gauge transformation. When one treats the gauge symmetry in the BV-BRST formalism, this means that observables are representatives of the cohomology of the BRST operator at zero ghost number~\cite{barnichetal2000}; for gauge theories where the gauge symmetry is an internal symmetry (such as Yang-Mills theories) one can then obtain a full classification of all local elements of this cohomology~\cite{barnichetal2000}, i.e., of all local observables. However, not all gauge transformations are pure redundancies: so-called large gauge transformations, which do not vanish at infinity or at the spacetime boundary, correspond to changes in the physical state of the system (to the addition of ``soft photons'' or ``soft gravitons''~\cite{weinberg1964,hemitraporfyriadisstrominger2014,kapeclysovpasterskistrominger2015,herdegen2017,condemao2017,hamadaseoshiu2017}), and have corresponding conserved charges~\cite{herdegen1995,barnichtroessaert2011,flanagannichols2017,comperelong2016,hawkingperrystrominger2017}. Therefore, we only demand invariance under gauge transformations which are different from zero only in a finite region in the interior of the spacetime.
\item \textbf{Quasi-locality:} It is of course well known that there are no local observables in a generally covariant theory, except at linear order in perturbations around a given background where one can, with some effort, also find a complete set of local gauge-invariant observables~\cite{froebhackhiguchi2017,froebhackkhavkine2018,khavkine2018}. In fact, this can be understood as the analogue of the non-existence of local charged fields in quantum electrodynamics: because of Gau{\ss}' law, charged fields must be dressed with a photon cloud which extends to infinity~\cite{kibble1968,kulishfaddeev1970,steinmann1984,baganlavellemcmullan2000a,baganlavellemcmullan2000b,mitraratabolesharatchandra2006}. However, in a perturbative treatment these non-localities only show up at higher orders, and we thus require that to lowest order in perturbation theory (and possibly also at first order) one recovers a strictly local observable. Moreover, at higher orders one might require the non-locality to be restricted in a suitable sense; for example, it seems reasonable to demand that the support of the observable is restricted to the past light cone, to avoid influences from processes happening at arbitrarily far spacelike separations (a form of undesirable ``action-at-a-distance'').
\end{enumerate}

The first criterion rules out observables like the one proposed by Tsamis and Woodard~\cite{tsamiswoodard2013} measuring the expansion rate in a de~Sitter background, which is only invariant under purely temporal gauge transformations. Such observables might be useful in more restricted contexts, and in fact their definition is motivated by the fact that the background spacetime is spatially homogeneous, such that for an also spatially homogeneous state (like the Bunch--Davies vacuum) non-invariance under spatial gauge transformations \emph{should} not matter (whether this is true in practice needs to be checked carefully). However, they are certainly not useful to study spatially inhomogeneous quantum states (e.g., states which are excited with respect to the Bunch--Davies vacuum), or spacetimes with small inhomgeneities or anisotropies in the background, and therefore neither can be used to compare quantum effects for such geometries and homogeneous ones. Other observables, such as the spatial averages proposed by Gasperini, Marozzi and Veneziano~\cite{gasperinimarozziveneziano2009,marozzi2011} that are invariant in the limit where the spatial averaging is performed on a full hypersurface extending to infinity, fall short of the second criterion. Moreover, they are again unsuitable for a generalisation to spatially inhomogeneous situations, and while their motivation is that taking a quantum expectation value in a homogeneous state effectively performs a spatial average, this is a property of the state and should not be incorporated into the observable itself.

One way to construct observables that satisfy both of the above criteria is the generalisation of the QED approach, ``dressing'' bare field operators with a graviton cloud~\cite{waresaotomeakhoury2013,donnellygiddings2015,donnellygiddings2016}. This seems especially suited for describing physical particles carrying their own gravitational field, which must be included to obtain a gauge-invariant description. Another proposal, suitable for more general constructions, was recently made by Brunetti et al.~\cite{brunettietal2016} and generalised by Fr{\"o}b and Lima~\cite{froeb2018,froeblima2018}. This proposal describes observables in a physical (field-dependent) coordinate system, and has the added advantage that its non-localities are causal, i.e., they are restricted to the past light cone. In the next section~\ref{sec_invobs}, we review this proposal both for a general background spacetime and for the special case of single-field inflation. In section~\ref{sec_hubble}, we construct an observable corresponding to the local Hubble rate in single-field inflation (or more generally Friedmann--Lema{\^\i}tre--Robertson--Walker cosmologies with an additional scalar degree of freedom), and calculate its expectation value to one-loop order for spacetimes of constant deceleration, which includes matter- and radiation-dominated expansion, and slow-roll inflation with vanishing second slow-roll parameter. We conclude in section~\ref{sec_results} with a discussion of the results and an outlook on future work. We use the `+++' convention of~\cite{mtw_book}, and set $c = \hbar = 1$ and $\kappa^2 = 16 \pi G_\text{N}$.

\section{Invariant observables}
\label{sec_invobs}

\subsection{General construction}
\label{sec_invobs_general}

The gauge-invariant observables that were considered by Brunetti et al.~\cite{brunettietal2016} and Fr{\"o}b and Lima~\cite{froeb2018,froeblima2018} belong to the class of relational observables. These are obtained by considering the field operator not at a point of the background spacetime, but instead at a point where another field has a given value~\cite{giddingsmarolfhartle2006,khavkine2015,marolf2015,brunettifredenhagenrejzner2016}. Relational observables were already studied long time ago~\cite{geheniaudebever1956a,geheniau1956,debever1956a,debever1956b,geheniaudebever1956b,komar1958,bergmannkomar1960,bergmann1961} (see~\cite{tambornino2012} for a recent review). In general, their construction involves scalars constructed from various fields, which are taken as configuration-dependent coordinates $\tilde{X}^{(\alpha)}[\phi]$, and observables are obtained by evaluating operators at a point where these configuration-dependent coordinates take a fixed value. In perturbation theory, one therefore needs a sufficiently generic background spacetime where one can differentiate points by the background values of these scalars. This is obviously problematic for perturbations around highly symmetric spacetimes; one can of course add the necessary scalar fields by hand (e.g., the famous Brown-Kucha{\v r} dust~\cite{brownkuchar1995}), but this changes the physical content. An illuminating example of this fact is given by the works by Giesel et al.~\cite{gieseletal2007,gieseletal2010}, where one can explicitly see the physical change in observables effected by the dust.

A way out of this dilemma is given by constructing the scalars (in perturbation theory) as solutions of a scalar differential equation, which is fulfilled in the background spacetime. For perturbations around Minkowski spacetime in Cartesian coordinates~\cite{froeb2018}, or more generally around arbitrary spacetimes in harmonic coordinates, we simply impose
\begin{equation}
\label{sec_invobs_general_harmonic}
\tilde{\nabla}^2 \tilde{X}^{(\alpha)}[\tilde{g}] = 0 \eqend{,}
\end{equation}
where $\tilde{\nabla}_\mu$ is the covariant derivative associated to the perturbed metric $\tilde{g}_{\mu\nu} = g_{\mu\nu} + \kappa h_{\mu\nu}$, with the background metric $g_{\mu\nu}$ and the perturbation $h_{\mu\nu}$. Note that the coordinates should be thought of as scalars, which is why the $\alpha$ index is enclosed in parentheses, and $\tilde{\nabla}^2$ is consequently the scalar d'Alembertian. Since the Cartesian coordinates in flat space are harmonic, this equation is fulfilled in the background spacetime, and one can then determine the $\tilde{X}^{(\alpha)}$ order by order in perturbation theory. Concretely, we expand them according to
\begin{equation}
\label{sec_invobs_general_coordpert}
\tilde{X}^{(\alpha)}[\tilde{g}] = x^\alpha + \sum_{k=1}^\infty \kappa^k X_{(k)}^{(\alpha)}[g,h] \eqend{,}
\end{equation}
where $X_k^{(\alpha)}$ contains $k$ powers of the perturbation $h_{\mu\nu}$. We also expand the inverse metric and the Christoffel symbols
\begin{equations}
\tilde{g}^{\mu\nu} &= g^{\mu\nu} + \sum_{k=1}^\infty \kappa^2 \delta g_{(k)}^{\mu\nu} \eqend{,} \\
\tilde{g}^{\mu\nu} \tilde{\Gamma}^\rho_{\mu\nu} &= g^{\mu\nu} \Gamma^\rho_{\mu\nu} + \sum_{k=1}^\infty \kappa^k \delta \Gamma_{(k)}^\rho \eqend{,}
\end{equations}
and impose equation~\eqref{sec_invobs_general_harmonic} at each order $k$. At zeroth order, the condition that the background coordinates be harmonic is $\nabla^2 x^\alpha = 0$, and at order $k \geq 1$ we obtain
\begin{equation}
\nabla^2 X_{(k)}^{(\alpha)} = \delta \Gamma_{(k)}^\alpha - \sum_{\ell=1}^{k-1} \left( \delta g_{(k-\ell)}^{\mu\nu} \partial_\mu \partial_\nu - \delta \Gamma_{(k-\ell)}^\rho \partial_\rho \right) X_{(\ell)}^{(\alpha)} \eqend{.}
\end{equation}

Choosing a Green's function $G(x,y)$ for the background d'Alembertian that fulfils
\begin{equation}
\nabla^2 G(x,y) = \frac{\delta^n(x-y)}{\sqrt{-g(x)}} \eqend{,}
\end{equation}
we can thus construct the $X_k^{(\alpha)}$ recursively:
\begin{equation}
\label{sec_invobs_general_xksol}
X_{(k)}^{(\alpha)}(x) = - \int G(x,y) \left[ \delta \Gamma_{(k)}^\alpha + \sum_{\ell=1}^{k-1} \left( \delta g_{(k-\ell)}^{\mu\nu} \partial_\mu \partial_\nu - \delta \Gamma_{(k-\ell)}^\rho \partial_\rho \right) X_{(\ell)}^{(\alpha)} \right](y) \sqrt{-g(y)} \total^n y \eqend{.}
\end{equation}
The non-locality inherent in these scalars obviously depends on the choice of Green's function. To ensure a causal evolution of the observables constructed with these scalars, one needs the retarded Green's function $G_\text{ret}$, for which the $\tilde{X}^{(\alpha)}$ reduce to the background coordinates $x^\alpha$ at past infinity. Whether such a choice is possible depends of course on the background spacetime. In many spacetimes of interest (including cosmological spacetimes), one can first choose an initial time $t_0$ and a Green's function $G_{t_0}$ such that
\begin{equation}
X^{(\alpha)}(t_0,\vec{x}) = x^\alpha \eqend{,} \qquad \partial_t X^{(\alpha)}(t_0,\vec{x}) = 0 \eqend{,}
\end{equation}
and then send $t_0$ to past infinity in a slightly complex direction, the so-called ``$\mathi \epsilon$'' prescription: $t_0 \to -\infty(1 - \mathi \epsilon)$ with $\epsilon > 0$. In this way, interactions at early times are suppressed, and an asymptotic adiabatic vacuum state for the full interacting theory is selected, analogous to the standard flat-space vacuum (see for instance section 4.2 in~\cite{peskinschroeder}, and~\cite{frv2011a} for a more detailed treatment in de~Sitter space). Moreover, this guarantees that no extra terms arise from integration by parts, which is needed to verify that the $X^{(\alpha)}$ do indeed transform as scalars.

The invariant observables are then defined by evaluating them at the spacetime point $x^\alpha$ corresponding to holding the $\tilde{X}^{(\alpha)}$ fixed. We thus have to invert the relation~\eqref{sec_invobs_general_coordpert}, which can easily be done by writing
\begin{equation}
x^\alpha = \tilde{X}^{(\alpha)} - \sum_{k=1}^\infty \kappa^k X_{(k)}^{(\alpha)}(x)
\end{equation}
and replacing $x^\alpha$ in the $X_{(k)}^{(\alpha)}$ on the right-hand side, recursively to any desired order. For example, up to second order we obtain
\begin{splitequation}
\label{sec_invobs_general_xintildex}
x^\alpha &= \tilde{X}^{(\alpha)} - \kappa X_{(1)}^{(\alpha)}(x) - \kappa^2 X_{(2)}^{(\alpha)}(x) + \bigo{\kappa^3} \\
&= \tilde{X}^{(\alpha)} - \kappa X_{(1)}^{(\alpha)}\left( \tilde{X} - \kappa X_{(1)}(\tilde{X}) \right) - \kappa^2 X_{(2)}^{(\alpha)}(\tilde{X}) + \bigo{\kappa^3} \\
&= \tilde{X}^{(\alpha)} - \kappa X_{(1)}^{(\alpha)} + \kappa^2 X_{(1)}^{(\mu)} \partial_\mu X_{(1)}^{(\alpha)} - \kappa^2 X_{(2)}^{(\alpha)} + \bigo{\kappa^3} \eqend{,}
\end{splitequation}
where the terms in the last line are all evaluated at $\tilde{X}$. For a scalar field $S$, with a perturbative expansion $S = S_{(0)} + \kappa S_{(1)} + \kappa^2 S_{(2)} + \bigo{\kappa^3}$, the corresponding invariant observable $\mathcal{S}$ is given by
\begin{splitequation}
\label{sec_invobs_general_invscalar}
\mathcal{S} = S(x) \Big\rvert_{\tilde{X} = \text{const}} &= S_0 + \kappa S_{(1)} - \kappa X_{(1)}^{(\mu)} \partial_\mu S_{(0)} + \kappa^2 S_{(2)} - \kappa^2 X_{(1)}^{(\mu)} \partial_\mu S_{(1)} \\
&\quad+ \kappa^2 \left[ X_{(1)}^{(\nu)} \partial_\nu X_{(1)}^{(\mu)} - X_{(2)}^{(\mu)} \right] \partial_\mu S_{(0)} + \frac{1}{2} \kappa^2 X_{(1)}^{(\mu)} X_{(1)}^{(\nu)} \partial_\mu \partial_\nu S_{(0)} + \bigo{\kappa^3} \eqend{,}
\end{splitequation}
where all terms in the last two lines are again evaluated at $\tilde{X}$, and the $X^{(\mu)}_{(k)}$ are given by their definition~\eqref{sec_invobs_general_xksol}. However, now the $\tilde{X}$ are just labels for points (as the coordinates $x^\mu$ had been before), and in particular one must not replace them by their expansion~\eqref{sec_invobs_general_coordpert} as this would just give back in the original scalar field $S$. In fact, once the explicit expression~\eqref{sec_invobs_general_invscalar} has been obtained, one may just rename $\tilde{X}$ to be $x$ again. To obtain invariant higher-spin fields, one also needs to include the Jacobian from the coordinate transformation $x \to \tilde{X}$. For example
\begin{equation}
\label{sec_invobs_general_invvector}
\mathcal{V}^\mu = \frac{\partial \tilde{X}^{(\mu)}}{\partial x^\rho} V^\rho(x) = \left( \frac{\partial x^\rho}{\partial \tilde{X}^{(\mu)}} \right)^{-1} V^\rho(x) \eqend{,}
\end{equation}
where the derivative is taken of the relation~\eqref{sec_invobs_general_xintildex} (or the analogue expansion to higher order), defines an invariant observable $\mathcal{V}^\mu$ from a vector field $V^\mu$. Since by their very definition the $\tilde{X}^{(\mu)}$ transform as scalars, the coordinates $x^\alpha$ viewed as functionals of the $\tilde{X}^{(\mu)}$, obtained by inverting the corresponding relation~\eqref{sec_invobs_general_xintildex}, transform inversely to a scalar. Evaluating the field $S$ (or $V^\mu$) at the position $x$ and holding $\tilde{X}$ fixed the transformations cancel out, and the final observable is invariant. Said otherwise, the change $x^\mu \to \tilde{X}^{(\mu)}$ is a field-dependent diffeomorphism, which has the effect of compensating for the explicit gauge transformation of fields by including the transformation of the metric perturbation.

\subsection{Observables for inflationary spacetimes}
\label{sec_invobs_inflation}

In an inflationary spacetime, which in this article means a (spatially flat) Friedmann--Lema{\^\i}tre--Robertson--Walker spacetime with an additional scalar degree of freedom, other choices of configuration-dependent coordinates are available. In particular, the scalar field serves as a natural clock, assuming that its gradient is everywhere timelike on the background. For simplicity, we present in the following explicit expressions for single-field inflation where the scalar is the inflaton, but the results can be extended to the general case (e.g., for a fluid with given equation of state) without difficulty. We thus assume a background spacetime with metric
\begin{equation}
\label{background_metric}
\total s^2 = g_{\mu\nu} \total x^\mu \total x^\nu = a^2(\eta) \left( - \total \eta^2 + \total \vec{x}^2 \right) \eqend{,}
\end{equation}
where $\eta$ is conformal time and $a$ the scale factor. The inflaton field $\phi$ has everywhere timelike gradient, taken w.l.o.g. to be $\phi' < 0$, and the background spacetime satisfies the Einstein--Klein--Gordon equations with a scalar potential $V(\phi)$. This results in the Friedmann equations
\begin{equations}[sec_invobs_inflation_eombg]
\kappa^2 V(\phi) &= 2 (n-2) (n-1-\epsilon) H^2 \eqend{,} \\
\kappa^2 (\phi')^2 &= 2 (n-2) \epsilon H^2 a^2 \eqend{,}
\end{equations}
where a prime denotes a derivative with respect to conformal time.

From the scale factor, we define the Hubble parameter $H$ and the first and second slow-roll parameters $\epsilon$ and $\delta$ as
\begin{equation}
\label{H_and_epsilon_def}
H \equiv \frac{a'}{a^2} \eqend{,} \qquad \epsilon \equiv - \frac{H'}{H^2 a} \eqend{,} \qquad \delta \equiv \frac{\epsilon'}{2 H a \epsilon} \eqend{,}
\end{equation}
They are related to the widely used Hubble slow-roll parameters $\epsilon_H$ and $\eta_H$ as~\cite{liddleparsonsbarrow1994}
\begin{equation}
\epsilon = \epsilon_H \eqend{,} \qquad \delta = \epsilon_H - \eta_H \eqend{,}
\end{equation}
and for a spacetime of constant $\epsilon$ and thus $\delta = 0$ (which we will concentrate on later), we have the exact expressions
\begin{equation}
\label{sec_invobs_inflation_constepsha}
H = H_0 a^{-\epsilon} \eqend{,} \qquad a = \left[ - (1-\epsilon) H_0 \eta \right]^{- \frac{1}{1-\epsilon}} \eqend{,} \qquad \eta = - \frac{1}{(1-\epsilon) H a} \eqend{.}
\end{equation}
For $\epsilon \to 0$, we then recover de~Sitter space, while a matter-dominated universe has $\epsilon_\text{mat} = (n-1)/2$ and radiation domination is $\epsilon_\text{rad} = n/2$. Taking a time derivative of the second Friedmann equation, one obtains the background scalar field equation
\begin{equation}
\label{sec_invobs_scalareom}
\phi'' = H a (1-\epsilon+\delta) \phi' \eqend{,}
\end{equation}
which will be useful later on.

We add perturbations according to
\begin{equations}[sec_invobs_inflation_pert]
g_{\mu\nu} &\to \tilde{g}_{\mu\nu} = a^2 \left( \eta_{\mu\nu} + \kappa h_{\mu\nu} \right) \eqend{,} \\
\phi &\to \tilde{\phi} = \phi + \kappa \phi^{(1)} \eqend{,}
\end{equations}
and as explained before the perturbed inflaton field serves as a natural clock. That is, we define
\begin{equation}
\label{sec_invobs_inflation_x0def}
\tilde{X}^{(0)} = \eta(\tilde{\phi}) \eqend{,}
\end{equation}
where $\eta(\phi)$ is obtained by inverting the background relation $\phi(\eta)$, and in particular we obtain to second order
\begin{equations}[sec_invobs_inflation_x0_sol_secondorder]
\tilde{X}^{(0)}_{(0)}(x) &= \eta \eqend{,} \\
\tilde{X}^{(0)}_{(1)}(x) &= \frac{\partial \eta(\phi)}{\partial \phi} \phi^{(1)}(x) = \frac{\phi^{(1)}(x)}{\phi'} \eqend{,} \\
\tilde{X}^{(0)}_{(2)}(x) &= \frac{1}{2} \frac{\partial^2 \eta(\phi)}{\partial \phi^2} \left[ \phi^{(1)}(x) \right]^2 = - \frac{\phi''}{2 (\phi')^3} \left[ \phi^{(1)}(x) \right]^2 = - \frac{(1-\epsilon+\delta) H a}{2 (\phi')^2} \left[ \phi^{(1)}(x) \right]^2 \eqend{.}
\end{equations}
The spatial coordinates are determined as before by imposing that they are annihilated by the perturbed d'Alembertian~\eqref{sec_invobs_general_harmonic},
\begin{equation}
\label{sec_invobs_inflation_xidef}
\tilde{\nabla}^2 \tilde{X}^{(i)}[\tilde{g}] = 0 \eqend{.}
\end{equation}
Expanding this equation to first order in perturbations, we have to solve~\cite{froeblima2018}
\begin{equation}
\left[ \partial^2 - (n-2) H a \partial_\eta \right] \tilde{X}^{(i)}_{(1)} = \partial_\nu h^{i\nu} - \frac{1}{2} \partial^i h + (n-2) H a h^{0i} \eqend{,}
\end{equation}
which gives
\begin{equation}
\label{sec_invobs_inflation_xi_sol_firstorder}
\tilde{X}^{(i)}_{(1)}(x) = \int G^\text{ret}_\text{H}(x,x') a^{n-2}(x') \left[ \partial_\nu h^{i\nu} - \frac{1}{2} \partial^i h + (n-2) H a h^{0i} \right](x') \total^n x' \eqend{.}
\end{equation}
Here, $G^\text{ret}_\text{H}(x,x')$ is a scalar retarded Green's function defined in~\cite{froeblima2018}, satisfying
\begin{equation}
\left[ \partial^2 - (n-2) H a \partial_\eta \right] G^\text{ret}_\text{H}(x,x') = a^{2-n} \delta^n(x-x') \eqend{,}
\end{equation}
whose explicit form will be given later on in section~\ref{sec_hubble}. While the original proposal by Brunetti et al.~\cite{brunettietal2016} involved the perturbed (covariant) Laplacian on equal-inflaton hypersurfaces instead of the d'Alembertian, the choice~\eqref{sec_invobs_inflation_xidef} together with the retarded Green's function in the perturbative construction ensures that the observables defined using these coordinates are supported in the past light cone, i.e., their non-localities are causal~\cite{froeblima2018}.

In the perturbed background~\eqref{sec_invobs_inflation_pert}, an infinitesimal coordinate transformation $x^\mu \to x^\mu - \kappa \xi^\mu$ leads to the following gauge transformations for metric and inflaton perturbations:
\begin{equations}[sec_invobs_inflation_gaugetrafo]
\delta_\xi h_{\mu\nu} &= 2 \partial_{(\mu} \xi_{\nu)} - 2 H a \eta_{\mu\nu} \xi_0 + \kappa \left( \xi^\rho \partial_\rho h_{\mu\nu} + 2 h_{\rho(\mu} \partial_{\nu)} \xi^\rho - 2 H a h_{\mu\nu} \xi_0 \right) \eqend{,} \\
\delta_\xi \phi^{(1)} &= - \xi_0 \phi' + \kappa \xi^\rho \partial_\rho \phi^{(1)} \eqend{.}
\end{equations}
It is then straightforward to check that the $\tilde{X}^{(\mu)}$ transform as scalars: to first order we obtain from the explicit expressions~\eqref{sec_invobs_inflation_x0_sol_secondorder} and~\eqref{sec_invobs_inflation_xi_sol_firstorder}
\begin{equations}
\delta_\xi \tilde{X}^{(0)}_{(1)}(x) &= \frac{\delta_\xi \phi^{(1)}(x)}{\phi'} = - \xi_0 + \bigo{\kappa} \eqend{,} \\
\begin{split}
\delta_\xi \tilde{X}^{(i)}_{(1)}(x) &= \int G^\text{ret}_\text{H}(x,x') a^{n-2}(x') \left[ \partial^2 - (n-2) (Ha)(x') \partial_{\eta'} \right] \xi^i(x') \total^n x' + \bigo{\kappa} \\
&= \xi^i + \bigo{\kappa} \eqend{,}
\end{split}
\end{equations}
which can be written as
\begin{equation}
\delta_\xi \tilde{X}^{(\mu)}_{(1)} = \xi^\mu + \bigo{\kappa} = \xi^\rho \partial_\rho x^\mu + \bigo{\kappa} \eqend{.}
\end{equation}
At second order, a similar but lengthy calculation gives
\begin{equation}
\delta_\xi \left[ \tilde{X}^{(\mu)}_{(1)} + \kappa \tilde{X}^{(\mu)}_{(2)} \right] = \xi^\rho \partial_\rho \left[ x^\mu + \kappa \tilde{X}^{(\mu)}_{(1)} + \bigo{\kappa^2} \right] \eqend{,}
\end{equation}
which is exactly the transformation of a scalar (up to that order). It then follows easily that the invariant scalar~\eqref{sec_invobs_general_invscalar} or vector observables~\eqref{sec_invobs_general_invvector} are indeed invariant; for example, to first order the scalar $S = S_{(0)} + \kappa S_{(1)} + \bigo{\kappa^2}$ transforms as
\begin{equation}
\delta_\xi S_{(0)} = 0 \eqend{,} \qquad \delta_\xi S_{(1)} = \xi^\rho \partial_\rho S_{(0)} + \bigo{\kappa} \eqend{,}
\end{equation}
and thus the invariant scalar~\eqref{sec_invobs_general_invscalar} does not transform:
\begin{splitequation}
\delta_\xi \mathcal{S} &= \delta_\xi \left[ S_0 + \kappa S_{(1)} - \kappa X_{(1)}^{(\mu)} \partial_\mu S_{(0)} + \bigo{\kappa^2} \right] \\
&= \kappa \delta_\xi S_{(1)} - \kappa \delta_\xi X_{(1)}^{(\mu)} \partial_\mu S_{(0)} + \bigo{\kappa^2} = 0 + \bigo{\kappa^2} \eqend{.}
\end{splitequation}
This invariance by construction persists also at higher orders (although its verification becomes increasily lengthy), and also for the invariant vector~\eqref{sec_invobs_general_invvector} (and other tensorial quantities defined analogously) we have
\begin{equation}
\delta_\xi V_{(0)}^\mu = 0 \eqend{,} \qquad \delta_\xi V_{(1)}^\mu = \xi^\rho \partial_\rho V_{(0)}^\mu - V_{(0)}^\rho \partial_\rho \xi^\mu + \bigo{\kappa} \eqend{,}
\end{equation}
and thus
\begin{equation}
\delta_\xi \mathcal{V}^\mu = \delta_\xi \left[ V_{(0)}^\mu + \kappa V_{(1)}^\mu - \kappa X_{(1)}^{(\rho)} \partial_\rho V_{(0)}^\mu + \kappa \partial_\rho X_{(1)}^{(\mu)} V_{(0)}^\rho + \bigo{\kappa^2} \right] = 0 + \bigo{\kappa^2} \eqend{.}
\end{equation}

For a given observable defined in this way, one can easily relate its linearised version with better known invariant observables in the linearised theory. For example, the invariant observable corresponding to the metric perturbation is
\begin{splitequation}
\label{sec_invobs_h00}
\mathcal{H}_{\mu\nu} &\equiv \frac{a^{-2}}{\kappa} \left[ \frac{\partial x^\rho}{\partial \tilde{X}^{(\mu)}} \frac{\partial x^\sigma}{\tilde{X}^{(\nu)}} \tilde{g}_{\rho\sigma}(x) - g_{\rho\sigma} \right] = h_{\mu\nu} - X_{(1)}^{(\rho)} \partial_\rho g_{\mu\nu} - 2 g_{\rho(\mu} \partial_{\nu)} X_{(1)}^{(\rho)} + \bigo{\kappa} \\
&= h_{\mu\nu} - 2 H a \eta_{\mu\nu} X_{(1)}^{(0)} - 2 \eta_{\rho(\mu} \partial_{\nu)} X_{(1)}^{(\rho)} + \bigo{\kappa} \eqend{,}
\end{splitequation}
where we used the background metric~\eqref{background_metric} and the definition of the Hubble parameter~\eqref{H_and_epsilon_def}. For its $00$ component, to which only the configuration-dependent time coordinate~\eqref{sec_invobs_inflation_x0_sol_secondorder} contributes, we obtain
\begin{equation}
\mathcal{H}_{00} = h_{00} + \frac{2}{\phi'} \left[ \partial_\eta + (\epsilon-\delta) H a \right] \phi^{(1)} + \bigo{\kappa} \eqend{,}
\end{equation}
where we used the background scalar equation of motion~\eqref{sec_invobs_scalareom}. When the constraint equations (i.e., the $00$ and the $0i$ components of the Einstein equations, which are elliptic equations) are fulfilled, one obtains in a somewhat lengthy but straightforward calculation that
\begin{equation}
\mathcal{H}_{00} = \frac{Q'}{H a}
\end{equation}
with the Sasaki-Mukhanov variable $Q$~\cite{mukhanovfeldmanbrandenberger1992}. However, while the Sasaki-Mukhanov variable is only invariant at linear order, $\mathcal{H}_{00}$ as defined by equation~\eqref{sec_invobs_h00} is invariant to all orders in perturbation theory. More explicit examples are given by Brunetti et al.~\cite{brunettietal2016}, where however an elliptic condition was imposed to determine the spatial invariant coordinates, which unlike the hyperbolic condition~\eqref{sec_invobs_inflation_xidef} imposed here suffers from causality violations~\cite{froeblima2018}. By such explicit calculations, one can easily compare the observables defined using this approach with other approaches to gauge-invariant variables in cosmology, for example the recent works by Giesel et al.~\cite{gieseletal2017,gieseletal2018} who also work in a relational setting. While the relational approach is of course not tied to any perturbative expansion, their explicit results only concern the linearised approximation, and it is not clear how higher-order approximations can be obtained. In fact, for invariant coordinates (``clocks'' in the terminology of~\cite{gieseletal2017,gieseletal2018}) determined without introducing extra matter fields, \cite{gieseletal2017} states explicitly: ``[...] it may be difficult to find appropriate clocks such that the power series [...] can be calculated in explicit form up to arbitrary high orders.'', and \cite{gieseletal2018} states: ``An open question that arises from our results is whether we can find non-linear geometrical clocks that reduce at the linear order to those we have identified here.'' In contrast, the invariant coordinates given in~\cite{brunettietal2016} and generalised in~\cite{froeblima2018} are given by the explicit formula~\eqref{sec_invobs_general_xksol} (and its analogue for other backgrounds) to arbitrary orders in perturbation theory. Moreover, the scalar--vector--tensor decomposition used in~\cite{gieseletal2017,gieseletal2018}, while standard in cosmology, introduces spatial non-localities and thus possible causality violations, which might cause problems with renormalisation in the quantum theory~\cite{brunettietal2016,froeblima2018}. Therefore, while the configuration-dependent time coordinate $\tilde{X}^{(0)}_{(1)}$ is equal to the ``uniform field gauge'' and the ``comoving gauge'' time variable used in~\cite{gieseletal2017,gieseletal2018} (up to a rescaling), the spatial coordinates are different. Linearised observables whose construction only involves $\tilde{X}^{(0)}_{(1)}$, such as $\mathcal{H}_{00}$ to linear order, thus agree in our approach and the one of~\cite{gieseletal2017,gieseletal2018} in either their ``uniform field gauge'' or ``comoving gauge'', while all others will disagree.

\section{The Hubble rate to second order}
\label{sec_hubble}

An important observable in cosmology is the local Hubble (or expansion) rate $H$, which measures the expansion of spacetime. In single-field inflation, it can be obtained from the divergence of the normalised gradient of the inflaton~\cite{geshnizjanibrandenberger2002}
\begin{equation}
H \equiv \frac{\nabla^\mu u_\mu}{n-1} \eqend{,} \qquad u_\mu \equiv \frac{\nabla_\mu \phi}{\sqrt{- \nabla^\mu \phi \nabla_\mu \phi}} \eqend{.}
\end{equation}
In the perturbed spacetime~\eqref{sec_invobs_inflation_pert}, we obtain
\begin{equation}
\tilde{H} = \frac{\tilde{\nabla}^\mu \tilde{u}_\mu}{n-1} = H + \kappa H^{(1)} + \kappa^2 H^{(2)} + \bigo{\kappa^3} \eqend{,}
\end{equation}
where the first-order correction $H^{(1)}$ is given by
\begin{equation}
\label{sec_hubble_rate_firstorder}
H^{(1)} = - \frac{\laplace \phi^{(1)}}{(n-1) a \phi'} + \frac{H}{2} h_{00} + \frac{1}{2 (n-1) a} \left( \partial_\eta h^k_k - 2 \partial^k h_{0k} \right) \eqend{,}
\end{equation}
and the second-order correction reads
\begin{splitequation}
\label{sec_hubble_rate_secondorder}
H^{(2)} &= - \frac{1}{4 (n-1) a} \left[ 2 h_{kl} \left( h^{kl}{}' - 2 \partial^k h_0^l \right) + 2 h_{0k} \left( \partial^k h^l_l - 2 \partial_l h^{kl} \right) - h_{00} \left( h^k_k{}' - 2 \partial^k h_{0k} \right) \right] \\
&\qquad+ \frac{3}{8} H h_{00} h_{00} - \frac{H}{2} h_{0k} h_0^k \\
&\quad+ \frac{1}{2 (n-1) a (\phi')^2} \left[ \partial_\eta \laplace \left( \phi^{(1)} \right)^2 - 2 \phi^{(1)} \partial_\eta \laplace \phi^{(1)} + (n-3+2\epsilon-2\delta) H a \partial^k \phi^{(1)} \partial_k \phi^{(1)} \right] \\
&\quad+ \frac{1}{2 (n-1) a \phi'} \left[ h_{00} \laplace \phi^{(1)} + 2 \partial_k h_{00} \partial^k \phi^{(1)} + 2 \partial_k \left( h^{kl} \partial_l \phi^{(1)} \right) - \partial_l h^k_k \partial^l \phi^{(1)} \right] \eqend{.}
\end{splitequation}

The invariant Hubble rate observable $\mathcal{H} = H + \kappa \mathcal{H}^{(1)} + \kappa^2 \mathcal{H}^{(2)} + \bigo{\kappa^3}$ is constructed according to the general procedure described in section~\ref{sec_invobs}, and we obtain
\begin{splitequation}
\mathcal{H}^{(1)} &= H^{(1)} - \tilde{X}^{(\mu)}_{(1)} \partial_\mu H = H^{(1)} + \frac{\epsilon H^2 a}{\phi'} \phi^{(1)} \\
&= \frac{(n-1) \epsilon H^2 a^2 \phi^{(1)} - \laplace \phi^{(1)}}{(n-1) a \phi'} + \frac{H}{2} h_{00} + \frac{1}{2 (n-1) a} \left( \partial_\eta h^k_k - 2 \partial^k h_{0k} \right)
\end{splitequation}
and
\begin{splitequation}
\mathcal{H}^{(2)} &= H^{(2)} - \tilde{X}^{(\mu)}_{(1)} \partial_\mu H^{(1)} - \tilde{X}^{(\mu)}_{(2)} \partial_\mu H + \tilde{X}^{(\mu)}_{(1)} \partial_\mu \tilde{X}^{(\nu)}_{(1)} \partial_\nu H + \frac{1}{2} \tilde{X}^{(\mu)}_{(1)} \tilde{X}^{(\nu)}_{(1)} \partial_\mu \partial_\nu H \\
&= H^{(2)} - \tilde{X}^{(\mu)}_{(1)} \partial_\mu H^{(1)} - \frac{H^2 a \epsilon}{\phi'} \tilde{X}^{(\mu)}_{(1)} \partial_\mu \phi^{(1)} + \frac{\epsilon-\delta}{2 (\phi')^2} H^3 a^2 \epsilon \left( \phi^{(1)} \right)^2 \eqend{,}
\end{splitequation}
where we have used the explicit expansion of the invariant time coordinate~\eqref{sec_invobs_inflation_x0_sol_secondorder}. We note that $\mathcal{H}$ measures the local expansion rate as seen by an observer that is co-moving with the coordinate system $\tilde{X}^{(\mu)}$; since $\tilde{X}^{(0)} = \eta(\tilde{\phi})$ this in particular means that the observer is co-moving with the inflaton. Using the gauge transformations of the metric and inflaton perturbations~\eqref{sec_invobs_inflation_gaugetrafo}, a long but straightforward calculation shows that $\mathcal{H}$ is indeed invariant, as it must be. One can thus calculate its expectation value in any gauge, and obviously the computation simplifies a lot in a gauge where $\tilde{X}^{(\mu)}_{(1)} = 0$. Using the explicit expansion of the invariant coordinates~\eqref{sec_invobs_inflation_x0_sol_secondorder}, \eqref{sec_invobs_inflation_xi_sol_firstorder} we see that this is the gauge where
\begin{equation}
\label{sec_hubble_gauge}
\phi^{(1)} = 0 = \partial_\nu h^{i\nu} - \frac{1}{2} \partial^i h + (n-2) H a h^{0i}
\end{equation}
exactly, i.e., also inside time-ordered products. This can be achieved by adding a Lagrange multiplier (auxiliary field) term to the action, and the corresponding propagator for the metric perturbation has been determined in~\cite{froeblima2018}. To second order, the interacting expectation value is given by
\begin{equation}
\label{sec_hubble_expectation}
\expect{ \mathcal{H}(x) } = H + \mathi \kappa^2 \expect{ H^{(1)}(x) S_\text{int}^{(1)} }_0 + \mathi \kappa^2 \expect{ H^{(1)}(x) S_\text{G,CT}^{(1)} }_0 + \kappa^2 \expect{ H^{(2)}(x) }_0 \eqend{,}
\end{equation}
where $S_\text{int}^{(1)}$ is the part of the full interaction (including the auxiliary field, gauge-fixing term and ghosts) linear in $\kappa$, $S_\text{G,CT}^{(1)}$ are the necessary gravitational counterterms, and $\expect{ \cdot }_0$ is the expectation value in the free theory in the particular gauge~\eqref{sec_hubble_gauge}.

Setting $\tilde{g}_{\mu\nu} = a^2 g_{\mu\nu}$, which results in
\begin{equations}[sec_hubble_conformal]
a^2 \tilde{R} &= R - 2 (n-1) a^{-1} \nabla^2 a - (n-1) (n-4) a^{-2} \nabla^\mu a \nabla_\mu a \eqend{,} \\
\sqrt{-\tilde{g}} &= a^n \sqrt{-g} \eqend{,}
\end{equations}
the gravitational and scalar field action
\begin{equation}
\label{sec_hubble_action}
S_\text{G} = \int \left[ \frac{1}{\kappa^2} \tilde{R} - \frac{1}{2} \tilde{\nabla}^\mu \tilde{\phi} \tilde{\nabla}_\mu \tilde{\phi} - \frac{1}{2} V(\tilde{\phi}) \right] \sqrt{-\tilde{g}} \total^n x
\end{equation}
reduces in the gauge where $\phi^{(1)} = 0$ to
\begin{equation}
S_\text{G} = \frac{1}{\kappa^2} \int \left[ a^{n-2} R \sqrt{-g} + (n-2) (n-1-\epsilon) H^2 a^n \left( g^{00} - 1 \right) \sqrt{-g} \right] \total^n x \eqend{,}
\end{equation}
after some integration by parts and use of the background scalar field equations~\eqref{sec_invobs_inflation_eombg}. Since for $g_{\mu\nu}$ the perturbative expansion is just $g_{\mu\nu} = \eta_{\mu\nu} + \kappa h_{\mu\nu}$ and consequently the Christoffel symbols are at least of first order in $h_{\mu\nu}$, this can be simplified further by expressing the Ricci scalar in terms of the Christoffel symbols and some further integration by parts. Using moreover that
\begin{splitequation}
&(n-1-\epsilon) H^2 a^n \sqrt{-g} \left( g^{00} - 1 \right) = \frac{1}{2} H a^{n-1} \sqrt{-g} g^{\mu\nu} g^{\rho\sigma} h_{0\sigma} \left( 2 \partial_\mu h_{\nu\rho} - \partial_\rho h_{\mu\nu} \right) \\
&\qquad+ H a^{n-1} \sqrt{-g} g^{\mu\nu} \left( - \partial_\mu h_{0\nu} + \partial_0 h_{\mu\nu} \right) + \text{total derivative} \eqend{,}
\end{splitequation}
we obtain in accordance with~\cite{iliopoulosetal1998}
\begin{splitequation}
S_\text{G} &= \frac{1}{4} \int g^{\mu\nu} g^{\alpha\beta} g^{\rho\sigma} \Big[ 2 \partial_\alpha h_{\mu\rho} \partial_\sigma h_{\nu\beta} - 2 \partial_\alpha h_{\mu\nu} \partial_\sigma h_{\rho\beta} - \partial_\alpha h_{\mu\rho} \partial_\beta h_{\nu\sigma} \\
&\qquad+ \partial_\alpha h_{\mu\nu} \partial_\beta h_{\rho\sigma} \Big] a^{n-2} \sqrt{-g} \total^n x + \frac{n-2}{2} \int g^{\mu\nu} g^{\rho\sigma} \Big[ h_{0\sigma} \partial_\rho h_{\mu\nu} \Big] H a^{n-1} \sqrt{-g} \total^n x \eqend{.}
\end{splitequation}
This form of the action makes it easy to extract the three-graviton interaction (linear in $\kappa$), and we obtain after renaming indices
\begin{splitequation}
\label{sec_hubble_interaction}
S_\text{G}^{(1)} &= \frac{1}{8} U^{\alpha\beta\gamma\delta\mu\nu\rho\sigma} \int a^{n-2} h_{\gamma\delta} \partial_\alpha h_{\mu\nu} \partial_\beta h_{\rho\sigma} \total^n x + \frac{n-2}{4} V^{\alpha\beta\mu\nu\rho\sigma} \int H a^{n-1} h_{\alpha\beta} h_{0\sigma} \partial_\rho h_{\mu\nu} \total^n x
\end{splitequation}
with
\begin{splitequation}
\label{sec_hubble_h2_tensoru}
U^{\alpha\beta\gamma\delta\mu\nu\rho\sigma} &= 2 \eta^{\mu\rho} \eta^{\alpha\sigma} \eta^{\nu\beta} \eta^{\gamma\delta} - 4 \eta^{\alpha\sigma} \eta^{\nu\beta} \eta^{\gamma\mu} \eta^{\delta\rho} - 4 \eta^{\mu\rho} \eta^{\nu\beta} \eta^{\alpha\gamma} \eta^{\sigma\delta} - 4 \eta^{\mu\rho} \eta^{\alpha\sigma} \eta^{\gamma\nu} \eta^{\delta\beta} \\
&\quad- 2 \eta^{\mu\nu} \eta^{\alpha\sigma} \eta^{\rho\beta} \eta^{\gamma\delta} + 4 \eta^{\alpha\sigma} \eta^{\rho\beta} \eta^{\gamma\mu} \eta^{\delta\nu} + 4 \eta^{\mu\nu} \eta^{\rho\beta} \eta^{\alpha\gamma} \eta^{\sigma\delta} + 4 \eta^{\mu\nu} \eta^{\alpha\sigma} \eta^{\gamma\rho} \eta^{\delta\beta} \\
&\quad- \eta^{\mu\rho} \eta^{\alpha\beta} \eta^{\nu\sigma} \eta^{\gamma\delta} + 2 \eta^{\alpha\beta} \eta^{\nu\sigma} \eta^{\gamma\mu} \eta^{\delta\rho} + 2 \eta^{\mu\rho} \eta^{\nu\sigma} \eta^{\alpha\gamma} \eta^{\beta\delta} + 2 \eta^{\mu\rho} \eta^{\alpha\beta} \eta^{\gamma\nu} \eta^{\delta\sigma} \\
&\quad+ \eta^{\mu\nu} \eta^{\alpha\beta} \eta^{\rho\sigma} \eta^{\gamma\delta} - 2 \eta^{\alpha\beta} \eta^{\rho\sigma} \eta^{\gamma\mu} \eta^{\delta\nu} - 2 \eta^{\mu\nu} \eta^{\rho\sigma} \eta^{\alpha\gamma} \eta^{\beta\delta} - 2 \eta^{\mu\nu} \eta^{\alpha\beta} \eta^{\gamma\rho} \eta^{\delta\sigma}
\end{splitequation}
and
\begin{equation}
\label{sec_hubble_h2_tensorv}
V^{\alpha\beta\mu\nu\rho\sigma} = \eta^{\mu\nu} \eta^{\rho\sigma} \eta^{\alpha\beta} - 2 \eta^{\rho\sigma} \eta^{\mu\alpha} \eta^{\nu\beta} - 2 \eta^{\mu\nu} \eta^{\alpha\rho} \eta^{\beta\sigma} \eqend{.}
\end{equation}

We also need to consider the gauge-fixing and ghost terms. The exact gauge~\eqref{sec_hubble_gauge} is imposed using a Lagrange multiplier (auxiliary field) $B_\mu$ in the form~\cite{froeblima2018}
\begin{equation}
S_\text{GF} = - \int \left[ a B_0 \phi^{(1)} - B_i \left( \partial_\nu h^{i\nu} - \frac{1}{2} \partial^i h + (n-2) H a h^{0i} \right) \right] a^{n-2} \total^n x \eqend{.}
\end{equation}
The corresponding ghost term is obtained by replacing $B_\mu$ by minus the antighost $\bar{c}_\mu$ and the metric and inflaton perturbations by their gauge transformations, with the gauge parameter replaced by the ghost:
\begin{equations}
\delta_c h_{\mu\nu} &= \partial_\mu c_\nu + \partial_\nu c_\mu - 2 H a \eta_{\mu\nu} c_0 + \kappa \left( c^\alpha \partial_\alpha h_{\mu\nu} + 2 h_{\alpha(\mu} \partial_{\nu)} c^\alpha - 2 H a h_{\mu\nu} c_0 \right) \eqend{,} \\
\delta_c \phi^{(1)} &= - c_0 \phi' \eqend{,}
\end{equations}
and we obtain
\begin{splitequation}
S_\text{GH} &= \int \left[ - a \phi' \bar{c}_0 c_0 - \bar{c}_i \left[ \partial^2 - (n-2) H a \partial_\eta \right] c^i \right] a^{n-2} \total^n x \\
&- \kappa \int \bar{c}_i \Big[ \partial_\nu \left( c^\alpha \partial_\alpha h^{i\nu} + 2 h^{\alpha(i} \partial^{\nu)} c_\alpha - 2 H a h^{i\nu} c_0 \right) - \frac{1}{2} \partial^i \left( c^\alpha \partial_\alpha h + 2 h_{\alpha\mu} \partial^\mu c^\alpha - 2 H a h c_0 \right) \\
&\qquad\quad+ (n-2) H a \left( c^\alpha \partial_\alpha h^{i0} + 2 h^{\alpha(i} \partial^{0)} c_\alpha - 2 H a h^{i0} c_0 \right) \Big] a^{n-2} \total^n x \eqend{.} \raisetag{1.9em}
\end{splitequation}
Since the free ghost action does not couple $\bar{c}_0$ and $c_i$ or $\bar{c}_i$ and $c_0$, and the ghost interaction term does not involve $\bar{c}_0$ at all, only the spatial (anti-)ghost will appear in loops. Moreover, since the gauge condition~\eqref{sec_hubble_gauge} is imposed exactly (even inside time-ordered products) we can use it to simplify the interaction terms. It follows that we can use the effective ghost action
\begin{splitequation}
\label{sec_hubble_ghost_action}
S_\text{GH,eff} &= - \int a^{n-2} \bar{c}_i \left[ \partial^2 - (n-2) H a \partial_\eta \right] c^i \total^n x \\
&\quad- \kappa \int \bar{c}_i \left[ \left( \partial_k h^{i\nu} + \partial^\nu h_k^i - \partial^i h_k^\nu \right) \partial_\nu c^k + h^{ki} \left[ \partial^2 - (n-2) H a \partial_\eta \right] c_k \right] a^{n-2} \total^n x \eqend{.}
\end{splitequation}
However, note that while the last term involves the equation of motion for the ghost one can not drop it from the interaction, since when acting on the time-ordered ghost propagator it will produce a $\delta$ distribution which in general will contribute to expectation values. From the free part we obtain the ghost propagator
\begin{equation}
\label{sec_hubble_ghost_propagator}
\expect{ c_i(x) \bar{c}_j(x') }_0 = \mathi \delta_{ij} G_\text{H}(x,x') \eqend{,}
\end{equation}
and the graviton propagator
\begin{equation}
\expect{ h_{\mu\nu}(x) h_{\rho\sigma}(x') } = \mathi G_{\mu\nu\rho\sigma}(x,x')
\end{equation}
had been determined in~\cite{froeblima2018}, with the result
\begin{equations}[sec_hubble_graviton_propagator]
G_{0000}(x,x') &= \frac{1}{(Ha)(\eta) (Ha)(\eta')} \partial_\eta \partial_{\eta'} G_\text{Q}(x,x') \eqend{,} \\
G_{000k}(x,x') &= \frac{\epsilon(\eta')}{2 (Ha)(\eta)} \partial_k D_\text{Q}(x,x') - \frac{1}{2 (Ha)(\eta) (Ha)(\eta')} \partial_\eta \partial_k G_\text{Q}(x,x') \eqend{,} \\
G_{00kl}(x,x') &= - \delta_{kl} \frac{1}{(Ha)(\eta)} \partial_\eta G_\text{Q}(x,x') \eqend{,} \\
\begin{split}
G_{0i0k}(x,x') &= \Pi_{ik} \left[ D_\text{H}(x,x') + D_2(x,x') \right] + \frac{\partial_i \partial_k}{\laplace} \bigg[ \frac{n-1}{2 (n-2)} D_\text{H}(x,x') + D_2(x,x') \\
&\qquad\quad+ \frac{(\epsilon H a)(\eta) \partial_\eta + (\epsilon H a)(\eta') \partial_{\eta'} - \laplace}{4 (H a)(\eta) (H a)(\eta')} G_\text{Q}(x,x') - \frac{\epsilon(\eta) \epsilon(\eta')}{4} D_\text{Q}(x,x') \bigg] \eqend{,}
\end{split} \\
\begin{split}
G_{0ikl}(x,x') &= - 2 \frac{\delta_{i(k} \partial_{l)}}{\laplace} \partial_\eta G_2(x,x') \\
&\quad- \delta_{kl} \frac{\partial_i}{\laplace} \left[ \frac{1}{n-2} \partial_\eta G_\text{H}(x,x') - \left[ \frac{\epsilon(\eta)}{2} \partial_\eta - \frac{\laplace}{2 (H a)(\eta)} \right] G_\text{Q}(x,x') \right] \eqend{,}
\end{split} \\
G_{ijkl}(x,x') &= \left( 2 \delta_{i(k} \delta_{l)j} - \frac{2}{n-2} \delta_{ij} \delta_{kl} \right) G_\text{H}(x,x') + \delta_{ij} \delta_{kl} G_\text{Q}(x,x') - 4 \frac{\partial_{(i} \delta_{j)(k} \partial_{l)}}{\laplace} G_2(x,x') \eqend{.}
\end{equations}
Here, $\Pi_{ij} \equiv \delta_{ij} - \partial_i \partial_j/\laplace$ is the transverse projector, and the various scalar propagators $G_{\text{H}/\text{Q}/2}$ and $D_{\text{H}/\text{Q}/2}$ are solutions of
\begin{equations}[sec_hubble_scalar_eom]
\left[ \partial^2 - (n-2) H a \partial_\eta \right] G_\text{H}(x,x') &= a^{2-n} \delta^n(x-x') \eqend{,} \\
\left[ \partial^2 - (n-2+2\delta) H a \partial_\eta \right] G_\text{Q}(x,x') &= \frac{2 a^{2-n}}{(n-2) \epsilon} \delta^n(x-x') \eqend{,} \\
\left[ \partial^2 - (n-2) H a \partial_\eta \right] G_2(x,x') &= \laplace G_\text{H}(x,x') \eqend{,}
\end{equations}
and
\begin{equations}
\laplace D_\text{H}(x,x') &= \partial_\eta \partial_{\eta'} G_\text{H}(x,x') - a^{2-n} \delta^n(x-x') \eqend{,} \\
\laplace D_\text{Q}(x,x') &= \partial_\eta \partial_{\eta'} G_\text{Q}(x,x') - \frac{2 a^{2-n}}{(n-2) \epsilon} \delta^n(x-x') \eqend{,} \\
\laplace D_2(x,x') &= \partial_\eta \partial_{\eta'} G_2(x,x') \eqend{.}
\end{equations}
All these propagators are time-ordered ones (i.e., Feynman propagators); for the corresponding Wightman functions the various $\delta$ distributions would be absent.

Lastly, we need the counterterms corresponding to renormalisations of the gravitational constant, the scalar field strength and the scalar potential. To one-loop order they can be obtained by expanding the action~\eqref{sec_hubble_action} to first order in the metric perturbation [using also the conformal transformation~\eqref{sec_hubble_conformal}], which in the gauge $\phi^{(1)} = 0$ gives
\begin{equation}
\label{sec_hubble_counterterm_action}
S_\text{G,CT}^{(1)} = (n-2) \int \left[ \epsilon \left( \delta_Z - \delta \kappa^2 \right) \left( h_{00} + \frac{1}{2} h \right) + \frac{1}{2} (n-1-\epsilon) \left( \delta \kappa^2 - \delta_V \right) h \right] H^2 a^n \total^n x \eqend{.}
\end{equation}
We note that one of the three renormalisations is redundant at this order, and we can set for example $\delta \kappa^2 = 0$. Moreover, in the de~Sitter limit $\epsilon \to 0$ the first combination vanishes, and only the renormalisation of the scalar potential (which then is proportional to the cosmological constant) remains. The computation of the Hubble rate expectation value~\eqref{sec_hubble_expectation} is now straightforward but somewhat tedious, and is done in the next two subsections.

\subsection{Pure second order term}
\label{sec_hubble_h2}

We calculate $\expect{ H^{(2)}(x) }_0$ by point-splitting the expression for $H^{(2)}$~\eqref{sec_hubble_rate_secondorder}, taking the expectation value and then the limit $x' \to x$. Using the propagator~\eqref{sec_hubble_graviton_propagator}, this results in
\begin{splitequation}
\label{sec_hubble_h2_hexpect_ing}
\expect{ H^{(2)}(x) }_0 &= \mathi \lim_{x' \to x} \bigg[ - \frac{n^2-2n-1}{4 (n-2) a} \left( \partial_\eta + \partial_{\eta'} \right) G_\text{H}(x,x') + \frac{1}{2 a} \left( \partial_\eta + \partial_{\eta'} \right) G_2(x,x') \\
&\qquad- \frac{1}{8 (n-1) H^2 a^3} \Big[ 2 (n-1) (H a)^2 \left( \partial_\eta + \partial_{\eta'} \right) + (n-1) H a \laplace \\
&\qquad\quad- (n-1) H a \partial_\eta \partial_{\eta'} + \left( \partial_\eta + \partial_{\eta'} \right) \laplace \Big] G_\text{Q}(x,x') - \frac{n-1}{2} H D_2(x,x') \\
&\qquad- H \frac{2 n^2 - 7 n + 7}{4 (n-2)} D_\text{H}(x,x') + \left( \frac{\epsilon}{4 (n-1) H a^2} \laplace + \frac{H \epsilon^2}{8} \right) D_\text{Q}(x,x') \bigg] \eqend{.}
\end{splitequation}
Note that since we are using time-ordered propagators, which are already symmetric in $x$ and $x'$, we do not need to explicitly symmetrise the point-split expression for $H^{(2)}$. Using the time-ordered propagators instead of the symmetrised two-point function has the advantage that renormalisation is simpler to perform, but gives otherwise an identical result.

For a spacetime with constant $\epsilon$ (and consequently $\delta = 0$), the scalar propagators simplify and we can express all of them in terms of $G_\text{H}$ and $D_\text{H}$. We have~\cite{froeblima2018}
\begin{equations}[sec_hubble_h2_props]
G_\text{Q}(x,x') &= \frac{2}{(n-2) \epsilon} G_\text{H}(x,x') \eqend{,} \\
G_2(x,x') &= - \frac{1}{2} \left( \eta \partial_\eta + \eta' \partial_{\eta'} - \frac{n-1-\epsilon}{1-\epsilon} \right) G_\text{H}(x,x') \eqend{,} \\
D_2(x,x') &= - \frac{1}{2} \left( \eta \partial_\eta + \eta' \partial_{\eta'} - \frac{n-3+\epsilon}{1-\epsilon} \right) D_\text{H}(x,x') \eqend{,} \\
D_\text{Q}(x,x') &= \frac{2}{(n-2) \epsilon} D_\text{H}(x,x') \eqend{,} \\
\left( \eta \partial_{\eta'} + \eta' \partial_\eta \right) G_\text{H}(x,x') &= \left( \eta \partial_\eta + \eta' \partial_{\eta'} - 2 \frac{n-2}{1-\epsilon} \right) D_\text{H}(x,x') \eqend{,}
\end{equations}
and using also the equations of motion~\eqref{sec_hubble_scalar_eom} and the relation~\eqref{sec_invobs_inflation_constepsha} it follows that
\begin{splitequation}
\label{sec_hubble_h2_hexpect_ing_consteps}
\expect{ H^{(2)}(x) }_0 &= \mathi \lim_{x' \to x} \bigg[ - \frac{2 + (2n^2-5n-1) \epsilon - (n^2-2n-1) \epsilon^2}{4 (n-2) a (1-\epsilon) \epsilon} \left( \partial_\eta + \partial_{\eta'} \right) G_\text{H}(x,x') \\
&\qquad- \frac{1 - (2n-3) \epsilon}{4 (n-2) H a^2 (1-\epsilon) \epsilon} \laplace G_\text{H}(x,x') \\
&\qquad+ \frac{n-1 + (2n^2-7n+7) \epsilon - 2 \epsilon^2}{4 (n-1) (n-2) \epsilon (1-\epsilon) H a^2} \partial_\eta \partial_{\eta'} G_\text{H}(x,x') \\
&\qquad- \frac{1}{4 (n-1) (n-2) \epsilon H^2 a^3} \left( \partial_\eta + \partial_{\eta'} \right) \laplace G_\text{H}(x,x') \\
&\qquad+ \frac{(n-3+\epsilon) (n^2-3n+3-\epsilon)}{4 (n-2) (1-\epsilon)} H D_\text{H}(x,x') \bigg] \eqend{.}
\end{splitequation}
We can now evaluate each of these terms individually. In Fourier space, we have (see, e.g.,~\cite{iliopoulosetal1998,froeblima2018})
\begin{equation}
\label{sec_hubble_h2_wightman_h_epsconst}
\tilde{G}^+_\text{H}(\eta,\eta',\vec{p}) = - \mathi (1-\epsilon)^\frac{(n-2) \epsilon}{1-\epsilon} H_0^\frac{n-2}{1-\epsilon} \frac{\pi}{4} (\eta\eta')^\mu \hankel1_\mu\left( -\abs{\vec{p}} \eta \right) \hankel2_\mu\left( -\abs{\vec{p}} \eta' \right)
\end{equation}
for the Wightman function, where the parameter $\mu$ reads
\begin{equation}
\label{sec_hubble_h2_mu_def}
\mu \equiv \frac{n-1-\epsilon}{2 (1-\epsilon)} \eqend{,}
\end{equation}
and
\begin{equation}
\label{sec_hubble_h2_feynman_h_epsconst}
\tilde{G}_\text{H}(\eta,\eta',\vec{p}) = \Theta(\eta-\eta') \tilde{G}^+_\text{H}(\eta,\eta',\vec{p}) + \Theta(\eta'-\eta) \tilde{G}^+_\text{H}(\eta',\eta,\vec{p}) \eqend{.}
\end{equation}
We see explicitly that the limit $\eta' \to \eta$ is the same from above and below, i.e., it does not matter whether we use the time-ordered or symmetrised two-point function in the expectation value~\eqref{sec_hubble_h2_hexpect_ing_consteps}. For the function $D^+_\text{H}$, we obtain using Hankel function identities~\cite{dlmf}
\begin{equation}
\label{wightman_dh_epsconst}
\tilde{D}^+_\text{H}(\eta,\eta',\vec{p}) = - \frac{\partial_\eta \partial_{\eta'}}{\vec{p}^2} \tilde{G}^+_\text{H}(\eta,\eta',\vec{p}) = \mathi (1-\epsilon)^\frac{(n-2) \epsilon}{1-\epsilon} H_0^\frac{n-2}{1-\epsilon} \frac{\pi}{4} (\eta\eta')^\mu \hankel1_{\mu-1}\left( -\abs{\vec{p}} \eta \right) \hankel2_{\mu-1}\left( -\abs{\vec{p}} \eta' \right) \eqend{,}
\end{equation}
and again $\tilde{D}_\text{H}(\eta,\eta',\vec{p}) = \Theta(\eta-\eta') \tilde{D}^+_\text{H}(\eta,\eta',\vec{p}) + \Theta(\eta'-\eta) \tilde{D}^+_\text{H}(\eta',\eta,\vec{p})$. Therefore, using again Hankel function identities~\cite{dlmf}, rescaling the integration variable and using the formul{\ae}~\eqref{sec_invobs_inflation_constepsha} and~\eqref{sec_hubble_h2_mu_def} for $\eta$, $H$, $a$ and $\mu$, we obtain
\begin{splitequation}
\label{sec_hubble_h2_detagh_limit}
\mathi \lim_{x' \to x} \left( \partial_\eta + \partial_{\eta'} \right) G_\text{H}(x,x') &= \mathi \lim_{x' \to x} \left( \partial_\eta + \partial_{\eta'} \right) \int \tilde{G}_\text{H}(\eta,\eta',\vec{p}) \mathe^{\mathi \vec{p} (\vec{x}-\vec{x}')} \frac{\total^{n-1} p}{(2\pi)^{n-1}} \\
&= - 2 (1-\epsilon)^\frac{(n-2) \epsilon}{1-\epsilon} H_0^\frac{n-2}{1-\epsilon} (-\eta)^{2\mu-n} J_{1,\mu,\mu-1} \\
&= - 2 (1-\epsilon) H^{n-1} a \, J_{1,\mu,\mu-1}
\end{splitequation}
with the dimensionless integral
\begin{splitequation}
J_{k,\alpha,\beta} &= \frac{\pi}{8} \int \abs{\vec{q}}^k \left[ \hankel1_\alpha\left( \abs{\vec{q}} \right) \hankel2_\beta\left( \abs{\vec{q}} \right) + \hankel2_\alpha\left( \abs{\vec{q}} \right) \hankel1_\beta\left( \abs{\vec{q}} \right) \right] \frac{\total^{n-1} q}{(2\pi)^{n-1}} \\
&= \frac{1}{2^n \pi^\frac{n-3}{2} \Gamma\left( \frac{n-1}{2} \right)} \Re \int_0^\infty \hankel1_\alpha\left( q \right) \hankel2_\beta\left( q \right) \, q^{k+n-2} \total q \eqend{.}
\end{splitequation}
In the same way, we obtain
\begin{equations}[sec_hubble_h2_moregh_limit]
\mathi \lim_{x' \to x} \laplace G_\text{H}(x,x') &= - (1-\epsilon)^2 H^n a^2 J_{2,\mu,\mu} \eqend{,} \\
\mathi \lim_{x' \to x} \partial_\eta \partial_{\eta'} G_\text{H}(x,x') &= (1-\epsilon)^2 H^n a^2 J_{2,\mu-1,\mu-1} \eqend{,} \\
\mathi \lim_{x' \to x} \left( \partial_\eta + \partial_{\eta'} \right) \laplace G_\text{H}(x,x') &= 2 (1-\epsilon)^3 H^{n+1} a^3 J_{3,\mu,\mu-1} \eqend{,} \\
\mathi \lim_{x' \to x} D_\text{H}(x,x') &= - H^{n-2} J_{0,\mu-1,\mu-1} \eqend{,} \\
\mathi \lim_{x' \to x} \left( \eta \partial_\eta + \eta' \partial_{\eta'} \right) D_\text{H}(x,x') &= 2 H^{n-2} \left( J_{1,\mu,\mu-1} - \frac{n-2}{1-\epsilon} J_{0,\mu-1,\mu-1} \right) \eqend{,}
\end{equations}
with the last following by using equation~\eqref{sec_hubble_h2_props} to express the derivatives of $D_\text{H}$ in terms of derivatives of $G_\text{H}$, and therefore
\begin{splitequation}
\label{sec_hubble_h2_hexpect_inj_consteps}
\expect{ H^{(2)}(x) }_0 &= H^{n-1} \bigg[ \frac{2 + (2n^2-5n-1) \epsilon - (n^2-2n-1) \epsilon^2}{2 (n-2) \epsilon} J_{1,\mu,\mu-1} \\
&\qquad+ \frac{n-1 + (2n^2-7n+7) \epsilon - 2 \epsilon^2}{4 (n-1) (n-2) \epsilon} (1-\epsilon) J_{2,\mu-1,\mu-1} \\
&\qquad+ \frac{1 - (2n-3) \epsilon}{4 (n-2) \epsilon} (1-\epsilon) J_{2,\mu,\mu} - \frac{(1-\epsilon)^3}{2 (n-1) (n-2) \epsilon} J_{3,\mu,\mu-1} \\
&\qquad- \frac{(n-3+\epsilon) (n^2-3n+3-\epsilon)}{4 (n-2) (1-\epsilon)} J_{0,\mu-1,\mu-1} \bigg] \eqend{.}
\end{splitequation}
By expanding the Hankel functions into Bessel functions, the $J$ integrals can be calculated analytically~\cite{dlmf} in terms of Gau{\ss}' hypergeometric function. Expressing its value at $z=1$ using $\Gamma$ functions, we obtain the result
\begin{splitequation}
\label{sec_hubble_h2_jintegral}
&J_{k,\mu-a,\mu-b} = (-1)^{a+b+k} 2^{k-1} \frac{\cos(\pi \mu) \Gamma\left( \frac{n+k-a+b-1}{2} \right) \Gamma\left( \frac{n+k+a-b-1}{2} \right)}{\pi^\frac{n+1}{2} \Gamma(n+k-1) \sin[ (n-4) \pi ] \Gamma\left( \frac{n-1}{2} \right)} \\
&\qquad\times \cos\left[ \frac{\pi}{2} (n+k+a+b) \right] \Gamma\left( \frac{n+k+a+b-1}{2} - \mu \right) \Gamma\left( \frac{n+k-a-b-1}{2} + \mu \right) \eqend{,}
\end{splitequation}
which is (as expected) divergent as $n \to 4$. Inserting this result we obtain
\begin{equation}
\label{sec_hubble_h2_hexpect_consteps}
\expect{ H^{(2)}(x) }_0 = - H^{n-1} C_2(n,\epsilon)
\end{equation}
with
\begin{splitequation}
C_2(n,\epsilon) &= \frac{\cos\left( \frac{n}{2} \pi \right) \cos(\pi \mu) \Gamma\left( \frac{n+1}{2} - \mu \right) \Gamma\left( \frac{n-3}{2} + \mu \right)}{2^{n+4} (n-2) \pi^\frac{n}{2} \Gamma\left( \frac{n+2}{2} \right) \sin[ (n-4) \pi ] \epsilon} \bigg[ 4 n (n^2+n-6) \\
&\quad+ 2 [ 8 + 28 n + (n+1) (2n-9) n^2 ] \epsilon + 8 (2n^2-4n-1) \epsilon^2 - n (n^2-4) \epsilon^3 \bigg] \eqend{,}
\end{splitequation}
which also is divergent as $n \to 4$. We now distinguish various subcases:
\begin{enumerate}
\item Matter domination, $\epsilon = (n-1)/2$ and $\mu = - (n-1)/[2 (n-3)]$:
\begin{equation}
C_2(n,\epsilon) = \frac{113}{96 \pi^2 (n-4)} + \bigo{(n-4)^0} \eqend{.}
\end{equation}
\item Radiation domination, $\epsilon = n/2$ and $\mu = - 1/2$:
\begin{equation}
C_2(n,\epsilon) = 0 \eqend{.}
\end{equation}
\item $n$-independent $\epsilon$ and $\mu_4 = (3-\epsilon)/[ 2(1-\epsilon) ]$:
\begin{splitequation}
\label{sec_hubble_h2_consteps_c2}
C_2(n,\epsilon) &= \frac{\cos(\pi \mu_4) \Gamma\left( \frac{5}{2} - \mu_4 \right) \Gamma\left( \frac{1}{2} + \mu_4 \right)}{128 \pi^3 (n-4) \epsilon} \left( 28 + 10 \epsilon + 15 \epsilon^2 - 6 \epsilon^3 \right) + \bigo{(n-4)^0} \\
&= \frac{14 + 33 \epsilon}{64 \pi^2 (n-4)} + \bigo{(n-4)^0} + \bigo{\epsilon^2} \eqend{.}
\end{splitequation}
\end{enumerate}

\subsection{Mixed first order -- counterterms}
\label{sec_hubble_ct}

For the counterterm contribution $\mathi \expect{ H^{(1)}(x) S_\text{G,CT}^{(1)} }_0$ to the expectation value of our observable, we calculate (taking already $\epsilon$ constant)
\begin{splitequation}
\label{sec_hubble_ct_h1sct}
\mathi \expect{ H^{(1)}(x) S_\text{G,CT}^{(1)} }_0 &= \frac{(n-2)}{4 (n-1) a} \left[ (n-1-\epsilon) \delta_V - \epsilon \delta_Z \right] \int F^k{}_k(x,x') (H^2 a^n)(x') \total^n x' \\
&\quad- \frac{(n-2) \epsilon}{2 (n-1) a} \delta_Z \int F_{00}(x,x') (H^2 a^n)(x') \total^n x'
\end{splitequation}
with
\begin{equation}
\label{sec_hubble_ct_fmunu}
F_{\mu\nu}(x,x') \equiv \partial_\eta G^k{}_{k\mu\nu}(x,x') - 2 \partial^k G_{0k\mu\nu}(x,x') + (n-1) H a G_{00\mu\nu}(x,x') \eqend{.}
\end{equation}
Using the propagator~\eqref{sec_hubble_graviton_propagator} and the constant-$\epsilon$ simplifications~\eqref{sec_hubble_h2_props}, we obtain
\begin{equations}
F^k{}_k(x,x') &= - \frac{2 (n-1)}{(n-2) \epsilon} \left[ \epsilon \partial_\eta - \frac{\laplace}{H a} \right] G_\text{H}(x,x') \eqend{,} \label{sec_hubble_ct_fkk} \\
F_{00}(x,x') &= - 2 \frac{(1-\epsilon) \eta'}{(n-2)} \laplace \left[ D_\text{H}(x,x') + \frac{1-\epsilon}{\epsilon} \eta \partial_{\eta'} G_\text{H}(x,x') \right] \label{sec_hubble_ct_f00} \eqend{.}
\end{equations}
Because of spatial homogeneity, the integrals in equation~\eqref{sec_hubble_ct_h1sct} will only depend on time. Since the spatial Laplacians in $F^k{}_k(x,x')$ and $F_{00}(x,x')$ act on $x$, we can take them out of the integral, and they will thus give no contribution to the result. It follows that we can also set $\delta_Z = 0$, and as in de~Sitter space~\cite{miaotsamiswoodard2017,tsamiswoodard2006} only the scalar potential (which there reduces to the cosmological constant) needs to be renormalised.

We thus obtain
\begin{equation}
\mathi \expect{ H^{(1)}(x) S_\text{G,CT}^{(1)} }_0 = - \frac{1}{2 a} (n-1-\epsilon) \delta_V K_2(\eta)
\end{equation}
with
\begin{equation}
K_m(\eta) \equiv \partial_\eta \int G_\text{H}(x,x') (H^m a^n)(\eta') \total^n x' \eqend{.}
\end{equation}
Since the integral $K_m(\eta)$ will also show up later, we calculate it separately. To ensure causality, the in-in (or Schwinger-Keldysh) formalism~\cite{schwinger1961,keldysh1965,chousuhaoyu1985} needs to be used to compute expectation values, which at one-loop order is equivalent to using the difference between the time-ordered (Feynman) propagator and negative frequency two-point (Wightman) function instead of just the time-ordered one~\cite{jordan1986,calzettahu1987}. For the integral $K_m(\eta)$ we then obtain a Heaviside $\Theta$ function which restricts the integration range to $\eta' \leq \eta$. Going to Fourier space, using the explicit expressions for the time-ordered $G_\text{H}$~\eqref{sec_hubble_h2_feynman_h_epsconst} and the two-point function~\eqref{sec_hubble_h2_wightman_h_epsconst} and performing the integral over the spatial coordinates, we obtain
\begin{splitequation}
\label{sec_hubble_ct_kmdef}
K_m(\eta) &= \partial_\eta \iint \Theta(\eta-\eta') \left[ \tilde{G}^+_\text{H}(\eta,\eta',\vec{p}) - \tilde{G}^+_\text{H}(\eta',\eta,\vec{p}) \right] \mathe^{\mathi \vec{p} (\vec{x}-\vec{x}')} \frac{\total^{n-1} p}{(2\pi)^{n-1}} (H^m a^n)(\eta') \total^n x' \\
&= \partial_\eta \int \Theta(\eta-\eta') \lim_{\vec{p} \to 0} \left[ \tilde{G}^+_\text{H}(\eta,\eta',\vec{p}) - \tilde{G}^+_\text{H}(\eta',\eta,\vec{p}) \right] (H^m a^n)(\eta') \total \eta' \eqend{.} \raisetag{2em}
\end{splitequation}
The combination in brackets has a finite limit as $\vec{p} \to 0$ (in any dimension $n$ and for any $\epsilon$)~\cite{dlmf}, and by also expressing $H$ and $a$ in terms of $\eta'$~\eqref{sec_invobs_inflation_constepsha} and taking the $\eta$ derivative inside the integral it follows that
\begin{splitequation}
\label{sec_hubble_ct_kmresult}
K_m(\eta) &= (1-\epsilon)^\frac{(n-2) \epsilon}{1-\epsilon} H_0^\frac{n-2}{1-\epsilon} \frac{1}{2 \mu} \partial_\eta \int \Theta(\eta-\eta') \left[ (-\eta)^{2\mu} - (-\eta')^{2\mu} \right] (H^m a^n)(\eta') \total \eta' \\
&= - (1-\epsilon)^\frac{- n + (n+m-2) \epsilon}{1-\epsilon} H_0^\frac{m-2}{1-\epsilon} (-\eta)^\frac{n-2}{1-\epsilon} \int_{-\infty}^\eta (-\eta')^{- \frac{n-m\epsilon}{1-\epsilon}} \total \eta' \\
&= - \frac{(1-\epsilon)^{-(n-2)}}{n-1-(m-1)\epsilon} H^{m-1} a \eqend{,}
\end{splitequation}
and we obtain
\begin{equation}
\label{sec_hubble_ct_result}
\mathi \expect{ H^{(1)}(x) S_\text{G,CT}^{(1)} }_0 = \frac{H}{2} (1-\epsilon)^{-(n-2)} \delta_V \eqend{.}
\end{equation}
Note that in principle one would have to use the $\mathi \epsilon$ prescription as explained in subsection~\ref{sec_invobs_general} to select the proper interacting vacuum state, calculating first the integral from $\eta_0$ to $\eta$ and then sending $\eta_0 \to -\infty(1-\mathi \tilde{\epsilon})$ with $\tilde{\epsilon} > 0$, and afterwards $\tilde{\epsilon} \to 0$. However, since the integral is already convergent without this prescription, it is unnecessary to use it explicitly, and we can set $\eta_0 = - \infty$ from the outset.

\subsection{Mixed first order -- interaction term}
\label{sec_hubble_h1s}

There are three contributions to $\mathi \expect{ H^{(1)}(x) S_\text{int}^{(1)} }_0$, which can be treated separately: one from the effective ghost action~\eqref{sec_hubble_ghost_action}, one from the interaction~\eqref{sec_hubble_interaction} involving the $U$ tensor, and one from the interaction~\eqref{sec_hubble_interaction} involving the $V$ tensor. For the ghost contribution, using the ghost~\eqref{sec_hubble_ghost_propagator} and graviton~\eqref{sec_hubble_graviton_propagator} propagators we obtain
\begin{splitequation}
&\mathi \expect{ H^{(1)}(x) S_\text{GH}^{(1)} }_0 = - \frac{\mathi}{2 (n-1) a} \int \partial_{x'}^\nu F(x,x') \left[ \lim_{u,v \to x'} \partial^u_\nu G_\text{H}(u,v) \right] a(x')^{n-2} \total^n x' \\
&\qquad- \frac{\mathi}{2 (n-1) a} \int F(x,x') \left[ \lim_{u,v \to x'} \left[ \partial^2 - (n-2) H a \partial_\eta \right] G_\text{H}(u,v) \right] a(x')^{n-2} \total^n x' \eqend{,}
\end{splitequation}
with $F_{\mu\nu}$ defined by equation~\eqref{sec_hubble_ct_fmunu}. For the terms in the second line, we obtain using the equation~\eqref{sec_hubble_scalar_eom} satisfied by $G_\text{H}$
\begin{equation}
\label{sec_hubble_h1s_ghost_eom}
\lim_{u,v \to x'} \left[ \partial^2 - (n-2) H a \partial_\eta \right] G_\text{H}(u,v) = \lim_{u,v \to x'} a^{2-n} \delta^n(u-v) = 0 \eqend{,}
\end{equation}
since $\delta^n(0) = 0$ in dimensional regularisation. For the terms in the first line, because of the spatial homogeneity we only obtain a non-vanishing result for $\nu = 0$, which for constant $\epsilon$ is given by ($1/2$ of) the result~\eqref{sec_hubble_h2_detagh_limit}. Furthermore, the spatial Laplacian in the result~\eqref{sec_hubble_ct_fkk} for the spatial trace $F^k{}_k(x,x')$ can be taken out of the integral since it acts on $x$, but since the result of the integral will only depend on time, it does not give a contribution. We thus obtain
\begin{splitequation}
\label{sec_hubble_h1s_ghost_contrib}
\mathi \expect{ H^{(1)}(x) S_\text{GH}^{(1)} }_0 &= \frac{1-\epsilon}{(n-2) a} J_{1,\mu,\mu-1} \int \partial_\eta \partial_{\eta'} G_\text{H}(x,x') [ (Ha)(\eta') ]^{n-1} \total^n x' \\
&= - \frac{(1-\epsilon)^2 (n-1)}{(n-2) a} J_{1,\mu,\mu-1} K_n(\eta) \eqend{,}
\end{splitequation}
where we have integrated the $\eta'$ derivative by parts, and the integral $K_n$ was defined by equation~\eqref{sec_hubble_ct_kmdef} and calculated in equation~\eqref{sec_hubble_ct_kmresult}. It follows that
\begin{equation}
\label{sec_hubble_h1s_expect_ghost}
\mathi \expect{ H^{(1)}(x) S_\text{GH}^{(1)} }_0 = - H^{n-1} C_\text{GH}(n,\epsilon)
\end{equation}
with [using the result~\eqref{sec_hubble_h2_jintegral} for the $J$ integral]
\begin{splitequation}
C_\text{GH}(n,\epsilon) &= - \frac{1}{n-2} (1-\epsilon)^{- (n-3)} J_{1,\mu,\mu-1} \\
&= \frac{\cos\left( \frac{n}{2} \pi \right) \cos(\pi \mu) \Gamma\left( \frac{n+1}{2} - \mu \right) \Gamma\left( \frac{n-1}{2} + \mu \right)}{2^{n-1} \pi^\frac{n}{2} (n-2) \Gamma\left( \frac{n}{2} \right) \sin[ (n-4) \pi ] (1-\epsilon)^{n-3}} \eqend{.}
\end{splitequation}

For the contributions from the three-graviton interaction terms~\eqref{sec_hubble_interaction} we proceed analogously. For the terms involving the $V$ tensor, we have
\begin{splitequation}
\label{sec_hubble_h1s_intv_contrib}
&\mathi \expect{ H^{(1)}(x) S_\text{G,V}^{(1)} }_0 = \frac{- \mathi (n-2)}{8 (n-1) a} V^{\alpha\beta\mu\nu\rho\sigma} \int F_{\alpha\beta}(x,x') \left[ \lim_{u,v \to x'} \partial^v_\rho G_{0\sigma\mu\nu}(u,v) \right] [ H a^{n-1} ](x') \total^n x' \\
&\qquad- \mathi \frac{n-2}{8 (n-1) a} V^{\alpha\beta\mu\nu\rho\sigma} \int F_{0\sigma}(x,x') \left[ \lim_{u,v \to x'} \partial^v_\rho G_{\alpha\beta\mu\nu}(u,v) \right] [ H a^{n-1} ](x') \total^n x' \\
&\qquad- \mathi \frac{n-2}{8 (n-1) a} V^{\alpha\beta\mu\nu\rho\sigma} \int \partial_{\rho'} F_{\mu\nu}(x,x') \left[ \lim_{u,v \to x'} G_{\alpha\beta0\sigma}(u,v) \right] [ H a^{n-1} ](x') \total^n x' \raisetag{2em}
\end{splitequation}
with $F_{\mu\nu}(x,x')$ defined by equation~\eqref{sec_hubble_ct_fmunu}, and $F_{00}$ given by equation~\eqref{sec_hubble_ct_f00}. Again we can take the spatial Laplacian out of the integral, which again will be a function of time only, such that all terms involving $F_{00}$ do not contribute. Similarly, also
\begin{splitequation}
\label{sec_hubble_h1s_f0i}
F_{0i}(x,x') &= \partial_i \left[ \frac{1+2n-n^2 + 2 (n-2) \epsilon - \epsilon^2}{2 (n-2) (1-\epsilon)} + \frac{1-\epsilon}{n-2} \eta \partial_\eta + \frac{n-1-\epsilon}{n-2} \eta' \partial_{\eta'} \right] D_\text{H}(x,x') \\
&\quad- \eta' \partial_i \left[ \partial_\eta - \frac{(1-\epsilon)^2}{(n-2) \epsilon} \eta \laplace \right] G_\text{H}(x,x') \raisetag{2em}
\end{splitequation}
is a total spatial derivative and all terms involving it do not contribute to the expectation value. For the same reason, the index $\rho$ in the last line of equation~\eqref{sec_hubble_h1s_intv_contrib} must be $0$, and the derivative can then be integrated by parts, resulting in
\begin{splitequation}
\label{sec_hubble_h1s_intv_contrib2}
\mathi \expect{ H^{(1)}(x) S_\text{G,V}^{(1)} }_0 = - \mathi \frac{n-2}{8 (n-1) a} &\int F_{ij}(x,x') \bigg[ [ H a^{n-1} ](x') V^{ij\mu\nu\rho\sigma} \lim_{u,v \to x'} \partial^v_\rho G_{0\sigma\mu\nu}(u,v) \\
&\quad- \partial_{\eta'} \left( [ H a^{n-1} ](x') V^{\alpha\beta ij0\sigma} \lim_{u,v \to x'} G_{\alpha\beta0\sigma}(u,v) \right) \bigg] \total^n x' \eqend{.}
\end{splitequation}
The evaluation of the coincidence limits is simplified by noting that because of spatial homogeneity only terms with an even number of spatial indices can contribute, which in the coincidence limit can only involve Kronecker $\delta$'s. Using the explicit form of the $V$ tensor~\eqref{sec_hubble_h2_tensorv}, we obtain
\begin{splitequation}
\label{sec_hubble_h1s_vg1}
&V^{\alpha\beta ij0\sigma} \lim_{u,v \to x'} G_{\alpha\beta0\sigma}(u,v) = \delta^{ij} \lim_{u,v \to x'} \left[ 2 G_0{}^k{}_{0k}(u,v) - \frac{n-3}{n-1} G^k{}_{k00}(u,v) - G_{0000}(u,v) \right] \\
&\quad= \frac{\delta^{ij}}{n-2} \left[ (n-1) (n-2) + 1-\epsilon \right] \lim_{x \to x'} \left( \eta \partial_\eta + \eta' \partial_{\eta'} - 2 \mu \right) D_\text{H}(x,x') \\
&\qquad+ \frac{\delta^{ij} (1-\epsilon)}{(n-2) \epsilon} \lim_{x \to x'} \left[ (1-\epsilon) \eta^2 \left( \laplace + 2 \partial_\eta \partial_{\eta'} \right) G_\text{H}(x,x') + 2 (n-3) \eta \partial_{\eta'} G_\text{H}(x,x') \right] \\
&\quad= - \mathi \frac{\delta^{ij}}{(n-2) (1-\epsilon) \epsilon} [ H(\eta') ]^{n-2} \Big[ (n^2-3n+3-\epsilon) \epsilon (n-3+\epsilon) J_{0,\mu-1,\mu-1} \\
&\qquad- 2 [ n-3 + (n^2-4n+6) \epsilon - \epsilon^2 ] (1-\epsilon) J_{1,\mu,\mu-1} + (1-\epsilon)^3 \left( J_{2,\mu,\mu} - 2 J_{2,\mu-1,\mu-1} \right) \Big] \eqend{,}
\end{splitequation}
where we used the limits~\eqref{sec_hubble_h2_detagh_limit} and~\eqref{sec_hubble_h2_moregh_limit} in the last step, and
\begin{splitequation}
\label{sec_hubble_h1s_vg2}
&V^{ij\mu\nu\rho\sigma} \lim_{u,v \to x'} \partial^v_\rho G_{0\sigma\mu\nu}(u,v) = \frac{\delta^{ij}}{n-1} \lim_{u,v \to x'} \Big[ (n-1) \partial^v_0 G_{0000}(u,v) - 2 (n-2) \partial^v_0 G_{00}{}^k{}_k(u,v) \\
&\hspace{16em}+ (n-3) \partial^k G_{0k00}(u,v) + (5-n) \partial^k G_{0k}{}^l{}_l(u,v) \Big] \\
&\quad= \frac{\delta^{ij}}{n-1} \lim_{x \to x'} \Big[ - \frac{2 (n-3) (n-1) - (n^2-2n-5) \epsilon - (n+1) \epsilon^2}{(n-2) \epsilon} \eta \partial_\eta \partial_{\eta'} G_\text{H}(x,x') \\
&\qquad\qquad+ (3n-5) \frac{(1-\epsilon)^2}{(n-2) \epsilon} \eta^2 \partial_\eta \laplace G_\text{H}(x,x') + (5-n) \frac{(n-1-\epsilon)}{(n-2) \epsilon} \eta \laplace G_\text{H}(x,x') \Big] \\
&\quad= \frac{- \mathi \, \delta^{ij}}{n-1} [ H^{n-1} a ](\eta') \Big[ \frac{2 (n-3) (n-1) - (n^2-2n-5) \epsilon - (n+1) \epsilon^2}{(n-2) \epsilon} (1-\epsilon) J_{2,\mu-1,\mu-1} \\
&\qquad\qquad+ (3n-5) \frac{(1-\epsilon)^3}{(n-2) \epsilon} J_{3,\mu,\mu-1} + (5-n) \frac{(n-1-\epsilon)}{(n-2) \epsilon} (1-\epsilon) J_{2,\mu,\mu} \Big] \eqend{,} \raisetag{2em}
\end{splitequation}
where we have also used the equation of motion of the scalar propagators~\eqref{sec_hubble_scalar_eom}. We note that the time-dependence of both expressions~\eqref{sec_hubble_h1s_vg1} and~\eqref{sec_hubble_h1s_vg2} is given by a simple power of the scale factor and the Hubble rate. This power is fixed by the behaviour of the scalar propagator in Fourier space~\eqref{sec_hubble_h2_wightman_h_epsconst} under the scaling $\eta \to \lambda \eta, \vec{p} \to \lambda^{-1} \vec{p}$, which leaves it invariant up to an overall factor, and at coincidence completely determines the time dependence. Since everything is proportional to $\delta^{ij}$, we see that only the spatial trace of $F_{ij}$ enters the expectation value~\eqref{sec_hubble_h1s_intv_contrib2}, which is given by~\eqref{sec_hubble_ct_fkk} (and again the spatial Laplacian does not contribute). It follows that
\begin{splitequation}
\label{sec_hubble_h1s_intv_contrib3}
\mathi \expect{ H^{(1)}(x) S_\text{G,V}^{(1)} }_0 &= \frac{1}{4 (n-1) (n-2) \epsilon a} K(\eta) \bigg[ (3n-5) (1-\epsilon)^3 J_{3,\mu,\mu-1} \\
&\qquad- (n-1)^2 (n^2-3n+3-\epsilon) \epsilon (n-3+\epsilon) J_{0,\mu-1,\mu-1} \\
&\qquad+ 2 (n-1)^2 [ n-3 + (n^2-4n+6) \epsilon - \epsilon^2 ] (1-\epsilon) J_{1,\mu,\mu-1} \\
&\qquad- [ 2 (n-3) (n-1) - (2n^2-3n-3) \epsilon + (n-1)^2 \epsilon^2 ] (1-\epsilon) J_{2,\mu,\mu} \\
&\hspace{-4em}+ [ 4 (n-2) (n-1) - (5n^2-10n-1) \epsilon + (2n^2-5n+1) \epsilon^2 ] (1-\epsilon) J_{2,\mu-1,\mu-1} \bigg] \\
&= - H^{n-1} C_\text{G,V}(n,\epsilon) \raisetag{1.2em}
\end{splitequation}
with
\begin{splitequation}
C_\text{G,V}(n,\epsilon) &= \frac{\cos\left( n \frac{\pi}{2} \right) \cos(\pi \mu) \Gamma\left( \frac{n+1}{2} - \mu \right) \Gamma\left( \frac{n-3}{2} + \mu \right)}{2^{n+5} (n-1) (n-2) \pi^\frac{n}{2} \Gamma\left( \frac{n+2}{2} \right) \sin[ (n-4) \pi ] \epsilon (1-\epsilon)^n} \\
&\quad\times \bigg[ 8 (n-2) (n-1) (5n-7) - 4 (2n^5-n^4+4n^3-68n^2+119n-42) \epsilon \\
&\qquad+ 4 (22n^4-74n^3+43n^2+47n-14) \epsilon^2 - n (53n^3-205n^2+222n-24) \epsilon^3 \\
&\qquad+ n (9n^3-37n^2+50n-16) \epsilon^4 \bigg] \eqend{.} \raisetag{2em}
\end{splitequation}

Lastly, for the terms involving the $U$ tensor we obtain
\begin{splitequation}
\label{sec_hubble_h1s_intu_contrib}
&\mathi \expect{ H^{(1)}(x) S_\text{G,U}^{(1)} }_0 = \frac{- \mathi}{16 (n-1) a} U^{\alpha\beta\gamma\delta\mu\nu\rho\sigma} \int a^{n-2}(x') F_{\gamma\delta}(x,x') \lim_{u,v \to x'} \partial^u_\alpha \partial^v_\beta G_{\mu\nu\rho\sigma}(u,v) \total^n x' \\
&\qquad- \mathi \frac{1}{16 (n-1) a} U^{\alpha\beta\gamma\delta\mu\nu\rho\sigma} \int a^{n-2}(x') \partial_{\alpha'} F_{\mu\nu}(x,x') \lim_{u,v \to x'} \partial^v_\beta G_{\gamma\delta\rho\sigma}(u,v) \total^n x' \\
&\qquad- \mathi \frac{1}{16 (n-1) a} U^{\alpha\beta\gamma\delta\mu\nu\rho\sigma} \int a^{n-2}(x') \partial_{\beta'} F_{\rho\sigma}(x,x') \lim_{u,v \to x'} \partial^v_\alpha G_{\gamma\delta\mu\nu}(u,v) \total^n x' \eqend{,} \raisetag{2.1em}
\end{splitequation}
and again only the spatial part of $F_{\mu\nu}$ contributes and the tensor structure of the coincidence limit of the last propagator can only contain Kronecker $\delta$'s, such that integrating by parts it follows that
\begin{splitequation}
\mathi \expect{ H^{(1)}(x) S_\text{G,U}^{(1)} }_0 = - \frac{\mathi}{16 (n-1)^2 a} &\int F(x,x') \delta_{ij} \bigg[ U^{\alpha\beta ij\mu\nu\rho\sigma} a^{n-2}(x') \lim_{u,v \to x'} \partial^u_\alpha \partial^v_\beta G_{\mu\nu\rho\sigma}(u,v) \\
&\quad- U^{0\beta\gamma\delta ij\rho\sigma} \partial_{\eta'} \left[ a^{n-2}(x') \lim_{u,v \to x'} \partial^v_\beta G_{\gamma\delta\rho\sigma}(u,v) \right] \\
&\quad- U^{\alpha0\gamma\delta\mu\nu ij} \partial_{\eta'} \left[ a^{n-2}(x') \lim_{u,v \to x'} \partial^v_\alpha G_{\gamma\delta\mu\nu}(u,v) \right] \bigg] \total^n x' \eqend{.}
\end{splitequation}
Invariance of the scalar propagators up to an overall power under the scaling $\eta \to \lambda \eta, \vec{x} \to \lambda \vec{x}$ mandates as before that the last two terms in brackets must be proportional to $(Ha)^{n-1}(\eta')$, which simplifies the calculation further since it follows that
\begin{splitequation}
\mathi \expect{ H^{(1)}(x) S_\text{G,U}^{(1)} }_0 &= - \frac{\mathi}{16 (n-1)^2 a} \int F(x,x') a^{n-2}(x') \delta_{ij} \lim_{u,v \to x'} \bigg[ U^{\alpha\beta ij\mu\nu\rho\sigma} \partial^u_\alpha \partial^v_\beta G_{\mu\nu\rho\sigma}(u,v) \\
&\qquad- (n-1) (1-\epsilon) H a \left( U^{0\beta\mu\nu ij\rho\sigma} + U^{\beta0\mu\nu\rho\sigma ij} \right) \partial^v_\beta G_{\mu\nu\rho\sigma}(u,v) \bigg] \total^n x' \eqend{.}
\end{splitequation}
For the coincidence limits, we obtain
\begin{splitequation}
&\delta_{ij} U^{\alpha\beta ij\mu\nu\rho\sigma} \lim_{u,v \to x'} \partial^u_\alpha \partial^v_\beta G_{\mu\nu\rho\sigma}(u,v) = \lim_{x \to x'} \bigg[ (5-n) \partial_\eta \partial_{\eta'} \left( G^k{}_k{}^l{}_l(x,x') - G^{kl}_{}{kl}(x,x') \right) \\
&\qquad\qquad+ 2 (n-3) \partial_\eta \partial^l \left( G_{00l0}(x,x') - G_{0l00}(x,x') \right) \\
&\qquad\qquad+ 2 (5-n) \partial_\eta \partial^l \left( G_{l0k}{}^k(x,x') + G^k{}_{kl0}(x,x') - 2 G^k{}_{lk0}(x,x') \right) \\
&\qquad\qquad+ 2 (5-n) \partial^k \partial^l \left[ G_{kl00}(x,x') - G_{0k0l}(x,x') + \delta_{kl} \left( G^i{}_{0i0}(x,x') - G^i{}_{i00}(x,x') \right) \right] \\
&\qquad\qquad+ (7-n) \partial^k \partial^l \left[ 2 G^i{}_{kil}(x,x') - 2 G^i{}_{ikl}(x,x') + \delta_{kl} \left( G^i{}_i{}^j{}_j(x,x') - G^{ij}_{}{ij}(x,x') \right) \right] \bigg] \\
&= \lim_{x \to x'} \bigg[ - 2 (n-3) \frac{(1-\epsilon)^2}{(n-2) \epsilon} \eta \eta' \left( \laplace + \partial_\eta \partial_{\eta'} \right) \laplace G_\text{H}(x,x') \\
&\qquad+ \left[ 2 (17-3n) - (n-3) (10-n) \frac{1-\epsilon}{(n-2) \epsilon} \right] \left( \eta \partial_\eta + \eta' \partial_{\eta'} \right) \laplace G_\text{H}(x,x') \\
&\qquad- \frac{1}{(n-2) \epsilon (1-\epsilon)} \Big[ 2 (n^3-11n^2+35n-37) - (3n^3-33n^2+104n-112) \epsilon \\
&\hspace{10em}- (n-3) (n^2-8n+6) \epsilon^2 \Big] \laplace G_\text{H}(x,x') \\
&\qquad+ \frac{1}{(n-2) \epsilon (1-\epsilon)} \Big[ 2 (5-n) (n-2) (n-1) + (-3n^3+25n^2-68n+62) \epsilon \\
&\hspace{10em}+ (n^3-5n^2+8n-8) \epsilon^2 - 2 (n-3) \epsilon^3 \Big] \partial_\eta \partial_{\eta'} G_\text{H}(x,x') \bigg] \\
&= \mathi [ H^n a^2 ](x') \bigg[ 2 (n-3) \frac{(1-\epsilon)^2}{(n-2) \epsilon} (1-\epsilon)^2 \left( J_{4,\mu,\mu} - J_{4,\mu-1,\mu-1} \right) \\
&\qquad\quad+ 2 (1-\epsilon)^2 \left[ 2 (17-3n) - (n-3) (10-n) \frac{1-\epsilon}{(n-2) \epsilon} \right] J_{3,\mu,\mu-1} \\
&\qquad\quad- \frac{(1-\epsilon)}{(n-2) \epsilon} \Big[ 2 (n^3-11n^2+35n-37) - (3n^3-33n^2+104n-112) \epsilon \\
&\hspace{10em} - (n-3) (n^2-8n+6) \epsilon^2 \Big] J_{2,\mu,\mu} \\
&\qquad\quad- \frac{(1-\epsilon)}{(n-2) \epsilon} \Big[ 2 (5-n) (n-2) (n-1) + (-3n^3+25n^2-68n+62) \epsilon \\
&\hspace{10em} + (n^3-5n^2+8n-8) \epsilon^2 - 2 (n-3) \epsilon^3 \Big] J_{2,\mu-1,\mu-1} \bigg]
\end{splitequation}
and
\begin{splitequation}
&\delta_{ij} \lim_{u,v \to x'} \left( U^{0\beta\mu\nu ij\rho\sigma} + U^{\beta0\mu\nu\rho\sigma ij} \right) \partial^v_\beta G_{\mu\nu\rho\sigma}(u,v) \\
&\quad= \lim_{x \to x'} \bigg[ \partial_\eta \left( 4 (n-3) G^{kl}{}_{kl}(x,x') - 2 (n-4) G^k{}_k{}^l{}_l(x,x') - 2 (n-2) G^k{}_{k00}(x,x') \right) \\
&\qquad\quad+ 2 (n-1) \partial^k G_{k000}(x,x') - 2 (n-1) \partial^k G_{k0l}{}^l(x,x') - 8 \partial^k G^l{}_{kl0}(x,x') \bigg] \\
&\quad= - \frac{2}{(n-2) \epsilon} \lim_{x \to x'} \bigg[ 2 (n-1) [ (n^2-7n+10) - (n-2) (n^2-3n-1) \epsilon ] \partial_\eta G_\text{H}(x,x') \\
&\hspace{8em}+ \left[ (n+1) (n-3) - (2n^3-10n^2+21n-21) \epsilon \right] \eta \laplace G_\text{H}(x,x') \\
&\hspace{8em}+ \left[ 2 (n-1) (n-2) - (2n^3-9n^2+18n-15) \epsilon + (n-1) \epsilon^2 \right] \eta \partial_\eta \partial_{\eta'} G_\text{H}(x,x') \\
&\hspace{8em}- (n-1) (1-\epsilon)^2 \eta \eta' \partial_{\eta'} \laplace G_\text{H}(x,x') \bigg] \\
&\quad= \mathi \frac{2 (1-\epsilon)}{(n-2) \epsilon} [ H^{n-1} a ](x') \bigg[ - 2 (n-1) \left[ (n^2-7n+10) - (n-2) (n^2-3n-1) \epsilon \right] J_{1,\mu,\mu-1} \\
&\hspace{8em}+ \left[ (n+1) (n-3) - (2n^3-10n^2+21n-21) \epsilon \right] J_{2,\mu,\mu} \\
&\hspace{8em}- \left[ 2 (n-1) (n-2) - (2n^3-9n^2+18n-15) \epsilon + (n-1) \epsilon^2 \right] J_{2,\mu-1,\mu-1} \\
&\hspace{8em}- (n-1) (1-\epsilon)^2 J_{3,\mu,\mu-1} \bigg] \eqend{,} \raisetag{1.8em}
\end{splitequation}
where we also needed the limits
\begin{equations}
\mathi \lim_{x' \to x} \laplace^2 G_\text{H}(x,x') &= (1-\epsilon)^4 H^{n+2} a^4 J_{4,\mu,\mu} \eqend{,} \\
\mathi \lim_{x' \to x} \partial_\eta \partial_{\eta'} \laplace G_\text{H}(x,x') &= - (1-\epsilon)^4 H^{n+2} a^4 J_{4,\mu-1,\mu-1} \eqend{.}
\end{equations}
It follows that
\begin{splitequation}
&\mathi \expect{ H^{(1)}(x) S_\text{G,U}^{(1)} }_0 = - \frac{1}{8 (n-2) (n-1) a} K(\eta) \bigg[ 2 (n-3) \frac{(1-\epsilon)^4}{(n-2) \epsilon} \left( J_{4,\mu,\mu} - J_{4,\mu-1,\mu-1} \right) \\
&\qquad+ 2 (1-\epsilon)^2 \left[ 2 (17-3n) - (n-3) (10-n) \frac{1-\epsilon}{(n-2) \epsilon} + (n-1)^2 \frac{(1-\epsilon)^2}{(n-2) \epsilon} \right] J_{3,\mu,\mu-1} \\
&\qquad- \frac{(1-\epsilon)}{(n-2) \epsilon} \Big[ 4 (n^3-7n^2+17n-17) + (-4n^4+19n^3-23n^2-18n+64) \epsilon \\
&\hspace{10em}+ (4n^4-25n^3+73n^2-114n+60) \epsilon^2 \Big] J_{2,\mu,\mu} \\
&\qquad+ \frac{(1-\epsilon)}{(n-2) \epsilon} \Big[ 2 (n-2) (n-1) (3n-7) + (-4n^4+21n^3-63n^2+114n-84) \epsilon \\
&\hspace{10em}+ (4n^4-23n^3+61n^2-78n+40) \epsilon^2 - 2 (n^3-3n+4) \epsilon^3 \Big] J_{2,\mu-1,\mu-1} \\
&\qquad+ 4 (n-1)^2 \frac{(1-\epsilon)^2}{(n-2) \epsilon} [ (n^2-7n+10) - (n-2) (n^2-3n-1) \epsilon ] J_{1,\mu,\mu-1} \bigg] \\
&= - H^{n-1} C_\text{G,U}(n,\epsilon) \raisetag{1.2em}
\end{splitequation}
with
\begin{splitequation}
C_\text{G,U}(n,\epsilon) &= \frac{\cos\left( n \frac{\pi}{2} \right) \cos(\pi \mu) \Gamma\left( \frac{n+1}{2} - \mu \right) \Gamma\left( \frac{n-1}{2} + \mu \right)}{2^{n+4} (n-1) (n-2)^2 \pi^\frac{n}{2} \Gamma\left( \frac{n+2}{2} \right) \sin[ (n-4) \pi ] \epsilon (1-\epsilon)^{n-1}} \\
&\quad\times \bigg[ - 8 (n-2) (n-1) (3n-2) + 2 (n-2) (5n^3+30n^2-73n+20) \epsilon \\
&\qquad\quad- (n-2) (23n^3+43n^2-158n+16) \epsilon^2 \\
&\qquad\quad+ (15n^4-17n^3-78n^2+104n+16) \epsilon^3 - 4 (n-1) n (n^2-2) \epsilon^4 \bigg] \eqend{.}
\end{splitequation}

For the various subcases, it follows that:
\begin{enumerate}
\item Matter domination, $\epsilon = (n-1)/2$ and $\mu = - (n-1)/[2 (n-3)]$:
\begin{equations}
C_\text{GH}(n,\epsilon) &= \frac{3}{4 \pi^2 (n-4)} + \bigo{(n-4)^0} \eqend{,} \\
C_\text{G,V}(n,\epsilon) &= - \frac{141}{8 \pi^2 (n-4)} + \bigo{(n-4)^0} \eqend{,} \\
C_\text{G,U}(n,\epsilon) &= - \frac{47}{8 \pi^2 (n-4)} + \bigo{(n-4)^0} \eqend{.}
\end{equations}
\item Radiation domination, $\epsilon = n/2$ and $\mu = - 1/2$:
\begin{equation}
C_\text{GH}(n,\epsilon) = C_\text{G,V}(n,\epsilon) = C_\text{G,U}(n,\epsilon) = 0 \eqend{.}
\end{equation}
\item $n$-independent $\epsilon$ and $\mu_4 = (3-\epsilon)/[ 2(1-\epsilon) ]$:
\begin{equations}[sec_hubble_h1s_consteps_cs]
\begin{split}
C_\text{GH}(n,\epsilon) &= \frac{\cos(\pi \mu_4) \Gamma\left( \frac{5}{2} - \mu_4 \right) \Gamma\left( \frac{3}{2} + \mu_4 \right)}{16 \pi^3 (n-4) (1-\epsilon)} + \bigo{(n-4)^0} \\
&= \frac{\epsilon}{8 \pi^2 (n-4)} + \bigo{(n-4)^0} + \bigo{\epsilon^2} \eqend{,}
\end{split} \\
\begin{split}
C_\text{G,V}(n,\epsilon) &= \frac{\cos(\pi \mu_4) \Gamma\left( \frac{5}{2} - \mu_4 \right) \Gamma\left( \frac{1}{2} + \mu_4 \right)}{768 \pi^3 (n-4) \epsilon (1-\epsilon)^4} \\
&\qquad\times (78 - 697 \epsilon + 879 \epsilon^2 - 488 \epsilon^3 + 84 \epsilon^4) + \bigo{(n-4)^0} \\
&= \frac{78 - 229 \epsilon}{768 \pi^2 (n-4)} + \bigo{(n-4)^0} + \bigo{\epsilon^2} \eqend{,}
\end{split} \\
\begin{split}
C_\text{G,U}(n,\epsilon) &= - \frac{\cos(\pi \mu_4) \Gamma\left( \frac{5}{2} - \mu_4 \right) \Gamma\left( \frac{3}{2} + \mu_4 \right)}{384 \pi^3 (n-4) \epsilon (1-\epsilon)^3} \\
&\qquad\times (30 - 132 \epsilon + 193 \epsilon^2 - 121 \epsilon^3 + 42 \epsilon^4) + \bigo{(n-4)^0} \\
&= - \frac{10 + 11 \epsilon}{64 \pi^2 (n-4)} + \bigo{(n-4)^0} + \bigo{\epsilon^2} \eqend{.}
\end{split}
\end{equations}
\end{enumerate}

\subsection{Renormalisation}
\label{sec_hubble_renorm}

To renormalise, we have to determine all operators with which $\mathcal{H}$ can mix. The suggestion of Miao et al.~\cite{miaotsamiswoodard2017} is
\begin{equations}[sec_hubble_renorm_miaocts]
\left( \tilde{R} \, \mathcal{H} \right)_{(0)} &= (n-1) (n-2\epsilon) H^3 \eqend{,} \\
\left( \mathcal{H}^3 \right)_{(0)} &= H^3 \eqend{,}
\end{equations}
and since $\epsilon = \text{const}$ these two are degenerate, i.e., our calculation cannot distinguish between the two operators. Moreover, we can construct a whole lot of other scalar operators (for example $\mathcal{H} \nabla^\mu \tilde{\phi} \nabla_\mu \tilde{\phi}$), which are all proportional to $H^3$ on the background, and can thus simply take $\left( \mathcal{H}^3 \right)_{(0)}$. For constant-$\epsilon$ spacetimes, we also need
\begin{equation}
\mathcal{H}_{(0)} = H \eqend{,}
\end{equation}
and it follows that
\begin{splitequation}
\expect{ \mathcal{H}_\text{ren} } &= \lim_{n \to 4} \left[ H - \kappa^2 H^{n-1} C(n,\epsilon) + \mu^{n-4} \alpha \kappa^2 \left( \mathcal{H}^3 \right)_{(0)} + \frac{H \kappa^2 \delta_V}{2 (1-\epsilon)^{n-2}} - \mu^{n-2} \beta \kappa^2 \mathcal{H}_{(0)} \right] \\
&= H - \kappa^2 H^3 \lim_{n \to 4} \left[ C(n,\epsilon) H^{n-4} - \alpha H_0^{n-4} - \frac{1}{2} (1-\epsilon)^{-(n-2)} H^{-2} \delta_V + \beta H_0^{n-2} H^{-2} \right] \eqend{,}
\end{splitequation}
with
\begin{equations}
C(n,\epsilon) &= C_1(n,\epsilon) + C_2(n,\epsilon) \eqend{,} \\
C_1(n,\epsilon) &= C_\text{GH}(n,\epsilon) + C_\text{G,V}(n,\epsilon) + C_\text{G,U}(n,\epsilon) \eqend{,}
\end{equations}
and where we have chosen the renormalisation scale $\mu$ to be equal to the Hubble rate $H_0$ at the initial time. It has been argued~\cite{tsamiswoodard2006,miaotsamiswoodard2017} that $\delta_V$ should be chosen such as to cancel the one-particle-irreducible contribution at the initial time where $a = 1$ and $H = H_0$, ensuring that $H_0$ is in effect the initial expansion rate at that time. This fixes $\delta_V$ to be
\begin{equation}
\delta_V = 2 (1-\epsilon)^{n-2} H_0^{n-2} C_1(n,\epsilon) \eqend{,}
\end{equation}
and to cancel the remaining divergence coming from $C_2(n,\epsilon)$ and from $C_1(n,\epsilon)$ at a time different from the initial time we take $\alpha = C(n,\epsilon)$ and $\beta = C_1(n,\epsilon)$. This gives
\begin{equation}
\expect{ \mathcal{H}_\text{ren} } = H - \kappa^2 H^3 \ln \frac{H}{H_0} \lim_{n \to 4} \left[ (n-4) C(n,\epsilon) \right] \eqend{,}
\end{equation}
such that in effect only the divergent part of $C(n,\epsilon)$ contributes. Since $H = H_0 a^{-\epsilon}$~\eqref{sec_invobs_inflation_constepsha}, we obtain finally
\begin{equation}
\label{sec_hubble_hren}
\expect{ \mathcal{H}_\text{ren} } = H \left[ 1 + \epsilon \kappa^2 H^2 \ln a \lim_{n \to 4} \left[ (n-4) C(n,\epsilon) \right] \right] \eqend{.}
\end{equation}

It may seem somewhat strange that we use the mixing with the operator $\mathcal{H}_{(0)}$ to cancel exactly the contribution coming from the counterterm $\delta_V$. However, the counterterm $\delta_V$ is already completely fixed by demanding that correlation functions with insertions of basic fields (the metric perturbation) are finite. Our observable $\mathcal{H}$ is a composite operator, for which additional renormalisation is necessary, and it is known (and can be rigorously proven~\cite{hollandswald2015}) that \emph{all} operators with the same quantum numbers and an equal or lower dimension are needed in general, with their coefficients given by background terms of the appropriate dimension. In our case, having the same quantum number restricts us to gauge-invariant operators, and the two necessary ones are $\left( \mathcal{H}^3 \right)_{(0)}$ whose coefficient (by dimensional analysis) must be proportional to $H_0^{n-4}$, and $\mathcal{H}_{(0)}$ whose coefficient must be proportional to $H_0^{n-2}$. This reasoning is in principle also valid in pure de~Sitter space, where however $H = H_0 = \text{const.}$, and the operators $\left( \mathcal{H}^3 \right)_{(0)}$ and $\mathcal{H}_{(0)}$ are degenerate (and in fact both just proportional to the unit operator).

\section{Results}
\label{sec_results}

For the various subcases, we obtain from equation~\eqref{sec_hubble_hren}:
\begin{enumerate}
\item Matter domination:
\begin{equation}
C(n,\epsilon) = - \frac{2071}{96 \pi^2 (n-4)} + \bigo{(n-4)^0}
\end{equation}
and
\begin{equation}
\label{sec_results_matter}
\expect{ \mathcal{H}_\text{ren} } = H \left[ 1 - \frac{2071}{64 \pi^2} \kappa^2 H^2 \ln a \right] \eqend{.}
\end{equation}
\item Radiation domination: $C(n,\epsilon) = 0$ and $\expect{ \mathcal{H}_\text{ren} } = H$.
\item $n$-independent $\epsilon$:
\begin{splitequation}
C(n,\epsilon) &= \frac{\cos(\pi \mu_4) \Gamma\left( \frac{5}{2} - \mu_4 \right) \Gamma\left( \frac{3}{2} + \mu_4 \right)}{768 \pi^3 (n-4) \epsilon (1-\epsilon)^3} \\
&\qquad\times (63 - 281 \epsilon + 90 \epsilon^2 - 22 \epsilon^3 + 108 \epsilon^4 - 162 \epsilon^5 + 36 \epsilon^6) + \bigo{(n-4)^0} \\
&= \frac{126 + 131 \epsilon}{768 \pi^2 (n-4)} + \bigo{(n-4)^0} \eqend{.}
\end{splitequation}
and
\begin{splitequation}
\expect{ \mathcal{H}_\text{ren} } &= H \left[ 1 + \epsilon \kappa^2 H^2 \ln a \lim_{n \to 4} \left[ (n-4) C(n,\epsilon) \right] \right] \\
&= H \left[ 1 + \frac{21}{128 \pi^2} \epsilon \kappa^2 H^2 \ln a + \bigo{\epsilon^2} \right] \eqend{.}
\end{splitequation}
\end{enumerate}
We see that for both matter domination and $n$-independent $\epsilon$ the invariant Hubble rate receives quantum corrections, while in the radiation-dominated case no corrections occur at one-loop order. Moreover, in general we obtain a secular effect, meaning that the quantum corrections grow in time, in our case as the logarithm of the scale factor. Therefore, at some point the quantum corrections become strong enough that perturbation theory breaks down, and one must employ, e.g., some form of resummation to obtain non-perturbative results. In particular, for small $\epsilon$ we can write
\begin{equation}
1 + \frac{21}{128 \pi^2} \epsilon \kappa^2 H^2 \ln a + \bigo{\epsilon^2} = a^{\frac{21}{128 \pi^2} \kappa^2 H_0^2 \epsilon} + \bigo{\epsilon^2}
\end{equation}
and therefore the last result can be written as
\begin{equation}
\label{sec_results_smalleps}
\expect{ \mathcal{H}_\text{ren} } = H(\hat{\epsilon}) + \bigo{\epsilon^2}
\end{equation}
with
\begin{equation}
\hat{\epsilon} = \epsilon \left( 1 - \frac{21}{128 \pi^2} \kappa^2 H_0^2 \right) \eqend{.}
\end{equation}
This means that quantum corrections move small constant $\epsilon$ spacetimes towards de~Sitter space where $\epsilon = 0$; clearly the one-loop correction also vanishes for pure de~Sitter space, which has also been found in other approaches~\cite{tsamiswoodard2006,miaotsamiswoodard2017}.\footnote{However, note that even in the $\epsilon \to 0$ limit the additional degree of freedom coming from inflaton perturbations remains. That is, the limit could be discontinuous and at higher loop orders one might obtain a different result from an analogous calculation in a pure de~Sitter background. I thank Albert Roura for discussions on this point.} In fact, we can analyse the various contributions~\eqref{sec_hubble_h2_consteps_c2} and~\eqref{sec_hubble_h1s_consteps_cs} to this result in more detail. Since $H = H_0 a^{-\epsilon}$~\eqref{sec_invobs_inflation_constepsha}, a positive correction to $\epsilon$ slows down the expansion of spacetime since $H$ gets smaller as $a$ grows, while a negative contribution accelerates it. Going back in the calculation, we thus see that a positive divergent part of $C(n,\epsilon)$ accelerates the expansion of spacetime and a negative divergent part slows it down. We see that the mutual attraction of gravitons, encoded in the interaction term (the sum $C_\text{G,U} + C_\text{G,V} + C_\text{GH}$), gives a negative divergent part and slows down the expansion, as it was proposed more than 25 years ago~\cite{ford1984,tsamiswoodard1993}. However, the contribution coming from the pure second-order term $C_2$ is positive and larger in magnitude, so that the overall effect is an accelerated expansion. This is seemingly in contradiction with the general analysis of Garriga and Tanaka~\cite{garrigatanaka2008}, who consider the expectation value of the Ricci scalar, averaged with a scalar window function $W(x)$ to make it gauge-invariant, and find a vanishing result. However, for their result to hold one needs a scalar $W(x)$ that is independent of the metric perturbations, which seems impossible to obtain without adding additional matter fields to the theory from which one could construct $W(x)$ --- and then one has changed the physical content of the theory under consideration.

That the corrections vanish for a radiation-dominated universe is perhaps surprising. However, it is in accordance with results obtained long ago~\cite{grishchuk1974} for the transverse traceless graviton modes only. Namely, for a scale factor which grows linearly with conformal time (which is the case for $\epsilon = 2$, the radiation domination), the equation of motion for transverse traceless graviton modes becomes conformal, and since the background is conformally flat no particle production takes place in this case~\cite{parker1969}.\footnote{I thank Bei-Lok Hu for bringing these references~\cite{grishchuk1974,parker1969} to my attention.} It is reassuring to see that this property persists for the full invariant observable $\mathcal{H}_\text{ren}$. That is, our full one-loop calculation of the expectation value of an invariant observable confirms that the physical process by which quantum corrections can change the acceleration of the background spacetime is, indeed, the production of low-energy gravitons which then mutually attract -- if no graviton production takes place, this process does not happen, and our observable does not introduce any spurious time dependence.

Since to our knowledge no other proposed observable in a single-field inflationary model fulfils the two criteria set out in the introduction, it is somewhat moot to compare the results~\eqref{sec_results_matter} and~\eqref{sec_results_smalleps} with existing ones. We thus restrict ourselves to the statement that existing calculations for comoving observers in pure de~Sitter space~\cite{miaotsamiswoodard2017} find a vanishing result, which is consistent with the $\epsilon \to 0$ limit of the small-$\epsilon$ result~\eqref{sec_results_smalleps}. The leading contributions from infrared modes in slow-roll spacetimes (again for comoving observers) also vanish~\cite{abramowoodard2002,geshnizjanibrandenberger2002}, but of course considering only long-wavelength modes can only account for a part of the complete result, which may or may not give the dominant contribution (the leading behaviour of scalar fields~\cite{starobinskyyokoyama1994,sasakinambunakao1988,finellietal2009,benekemoch2013,nacirmazzitellitrombetta2016} is correctly reproduced, but the treatment of gravitons is much more difficult~\cite{tsamiswoodard2005}).

It would be valuable to extend the above results to two-loop corrections, where also in pure de~Sitter space a non-trivial result is expected~\cite{miaotsamiswoodard2017}, and to general slow-roll spacetimes. In particular, the extension to a slow-roll spacetime with $\delta \neq 0$ would allow to differentiate between the two counterterms~\eqref{sec_hubble_renorm_miaocts} (and possibly others), showing which one mixes with the one-loop corrections to $\mathcal{H}$.

\ack
This work is part of a project that has received funding from the European Union's Horizon 2020 research and innovation programme under the Marie Sk{\l}odowska-Curie grant agreement No. 702750 ``QLO-QG''. I thank William C. C. Lima for collaboration in an early stage of the work, Bei-Lok Hu for references and comments, Richard Woodard for comments, and the anonymous referees for comments and reference suggestions.

\appendix

\providecommand\newblock{\ }
\bibliography{literature}

\end{document}